\pdfoutput=1             

\documentclass[preprint,12pt,authoryear]{elsarticle}

\makeatletter
\def\ps@pprintTitle{%
   \let\@oddhead\@empty
   \let\@evenhead\@empty
   \let\@oddfoot\@empty
   \let\@evenfoot\@oddfoot}
\makeatother

\usepackage{setspace} 
\usepackage{lineno}   
\usepackage[nohyperlinks]{acronym}
\usepackage[hidelinks]{hyperref}

\usepackage{wrapfig} 
\usepackage{lineno} 
\usepackage{appendix}

\usepackage{tikz}

\usetikzlibrary{shapes.geometric, arrows} 
\usepackage{pdfpages}
\usepackage[T1]{fontenc}
\usepackage[utf8]{inputenc}
\usepackage{lmodern}
\usepackage{amsmath,amssymb,mathtools}
\usepackage[capitalise, noabbrev]{cleveref} 
\usepackage{graphicx}
\usepackage{booktabs}
\usepackage{siunitx}
\usepackage{doi}
\usepackage{listings}
\usepackage{caption}
\usepackage{subcaption} 
\usepackage{natbib}
\usepackage{doi} 

\begin{document}

\doublespacing
\acresetall
\title{Trans-Arctic route feasibility on a pan-Arctic grid under bathymetric and sea-ice constraints}
\author[label1,label2]{Abdella Mohamed\corref{mycorrespondingauthor}}
\ead{mohamed.abdella@tum.de}

\author[label1]{Xiangyu Hu\corref{mycorrespondingauthor}}
\ead{xiangyu.hu@tum.de}

\cortext[mycorrespondingauthor]{Corresponding author}
\affiliation[label1]{organization={Technical University of Munich},
         addressline={Boltzmannstraße 15},
         postcode={85748},
         city={Garching},
         country={Germany}}
\affiliation[label2]{organization={Everllence (formerly: MAN Energy Solutions)},
         addressline={Stadtbachstraße 1},
         postcode={86153},
         city={Augsburg},
         country={Germany}}

\begin{frontmatter}

    \begin{abstract}

Climate-driven reductions in Arctic sea-ice extent have renewed interest in trans-Arctic shipping, yet adoption remains limited by questions of route feasibility, safety, and excess distance. Existing work often compares idealised great-circle shortcuts or uses detailed weather-routing systems that couple metocean forcing, sea ice and ship performance, leaving a gap for basin-scale diagnostics on realistic bathymetry and sea-ice fields. We develop an offline graph-based framework for trans-Arctic route feasibility on a $0.5^\circ$ pan-Arctic grid that combines General Bathymetric Chart of the Oceans (GEBCO)~2024 bathymetry with a summer 2018 Arctic sea-ice reanalysis from the Copernicus Marine Environment Monitoring Service (CMEMS). Bathymetry is regridded to construct sea-only and depth-feasibility masks, while daily sea-ice concentration (SIC) fields define ice-feasibility masks. An A* pathfinding algorithm is applied to a canonical Europe--Asia origin--destination pair to quantify route availability and route-length inflation relative to great-circle baselines.

Enforcing sea-only feasibility increases route length by about 10\% before depth and ice constraints are applied. At $0.5^\circ$ resolution, depth thresholds representative of under-keel clearance ($h_\mathrm{min}=20$--$50\,\mathrm{m}$) remove up to roughly 15\% of the sea mask but preserve a trans-Arctic connection for $h_\mathrm{min}=20\,\mathrm{m}$. 
Summer sea ice exerts a strong seasonal control: continuous ice-safe routes
emerge only from mid-August, with distances inflated by roughly 20–25
even in late summer only about 75\% of sea cells are simultaneously depth- and
ice-safe, leaving no continuous joint-safe trans-Arctic route in the tested
season. These diagnostics position the framework as a basin-scale screening tool for Arctic shipping and as a baseline for subsequent forecast-driven, multi-objective Arctic routing studies.
    \end{abstract}


\end{frontmatter}

\section*{Nomenclature}
\renewcommand{\baselinestretch}{0.75}\normalsize
\renewcommand{\aclabelfont}[1]{\textsc{\acsfont{#1}}}
\begin{acronym}[longest]

\acro{astar}[A*]{A-star shortest path algorithm}

\acro{cmems}[CMEMS]{Copernicus Marine Environment Monitoring Service}
\acro{cmip}[CMIP]{Coupled Model Intercomparison Project}
\acro{co2}[CO\textsubscript{2}]{Carbon dioxide}

\acro{gc}[GC]{Great-circle}
\acro{gebco}[GEBCO]{General Bathymetric Chart of the Oceans}

\acro{nep}[NEP]{Northeast Passage}
\acro{netcdf}[NetCDF]{Network Common Data Form}
\acro{nsr}[NSR]{Northern Sea Route}

\acro{od}[OD]{Origin--destination}

\acro{sic}[SIC]{Sea-ice concentration}
\acro{suez}[Suez]{Suez Canal Route}

\acro{wmo}[WMO]{World Meteorological Organization}

\end{acronym}
\renewcommand{\baselinestretch}{1}\normalsize

\section{Introduction}\label{sec:introduction}

Growing interest in decarbonisation, shifting trade patterns and declining Arctic sea-ice extent have renewed attention to high-latitude shipping corridors linking northern Europe and northeast Asia \citep{Stroeve2012,Notz2016}. Trans-Arctic options such as the Northern Sea Route (NSR) and related passages are often presented as 30--40\% ``shortcuts'' relative to traditional Suez route, with implied savings in time, fuel and \(\mathrm{CO_2}\) emissions \citep{Melia2016}. At the same time, regulatory uncertainty, sparse infrastructure and safety concerns have so far limited large-scale commercial uptake \citep{Stephenson2013,Chen2021TC}.

A growing body of research has examined different aspects of Arctic route feasibility and potential, but important gaps remain at basin scale under explicit bathymetry and sea-ice constraints. One strand uses climate model projections and simple sea-ice indices to map potential future Arctic navigability, typically in terms of seasonal windows or navigable fractions of predefined corridors \citep{Stephenson2013,Melia2016,Aksenov2017,Chen2021TC,Zhang2023JMSE}. These studies highlight the strong sensitivity of Arctic shipping potential to greenhouse-gas scenarios and ice thresholds, but usually do not construct explicit ship routes on realistic coastal and bathymetric grids. A second strand focuses on detailed techno-economic assessments along specific Arctic routes, combining assumed voyage distances with fuel, emissions and cost models to compare NSR and Suez itineraries under various ice and regulatory regimes \citep{Zhu2018,Pruyn2022}. A third strand embeds sea-ice and metocean information directly in operational weather-routing or ship-ice interaction tools, coupling environmental forcing with vessel performance and optimisation algorithms to produce case-specific routes or risk metrics \citep{Kuuliala2017,Lu2021,Liu2023}. While powerful, such systems are often complex and vessel-specific, making them less suited to transparent, basin-scale diagnostics.

Beyond these strands, many corridor-scale comparisons between Arctic and traditional Europe--Asia routes still rely on great-circle or waypoint-based sea-only distances to infer changes in voyage time, fuel consumption and \(\mathrm{CO_2}\) emissions between NSR, Suez and Cape of Good Hope baselines. In most such studies, route feasibility is handled implicitly through the placement of waypoints, without explicit bathymetric or sea-ice safety masks. From the perspective of end-users such as shipowners, cargo interests and policy analysts, there remains a need for simpler, physically grounded diagnostics that answer two prior questions: (i) does a continuous trans-Arctic connection exist at all under basic depth and ice criteria, and (ii) if so, what is the associated distance penalty relative to a great-circle lower bound?

This paper aims to bridge part of that gap by developing an offline, graph-based routing framework for historical trans-Arctic route diagnostics under explicit bathymetric and sea-ice constraints. We consider a pan-Arctic corridor from \(40^\circ\)N to \(85^\circ\)N and \(20^\circ\)W to \(180^\circ\)E discretised at \(0.5^\circ\) resolution, and combine GEBCO~2024 bathymetry with a daily Arctic sea-ice concentration (SIC) reanalysis from the Copernicus Marine Environment Monitoring Service (CMEMS) for the June--September 2018 season. A canonical Europe--Asia origin--destination pair is chosen to span the basin without tying the analysis to particular ports, in line with recent work emphasising generic corridor diagnostics over case-specific itineraries \citep{Zhang2023JMSE}. On this grid we construct sea-only, depth-constrained, ice-constrained and joint depth-plus-ice feasibility masks, and apply an A* pathfinding algorithm to quantify route availability, distance and excess length relative to great-circle baselines.

The analysis is intentionally restricted to static bathymetry and summer 2018 SIC fields, without metocean forcing, forecasts or vessel-specific performance models. This keeps the framework simple enough for exhaustive seasonal and threshold-sensitivity experiments, and positions the results as an envelope of what is theoretically possible under idealised operational behaviour. At the same time, the use of realistic coastlines, modern bathymetry and regridded sea-ice fields ensures that the diagnostics reflect key physical constraints on large-scale Arctic shipping rather than purely geometric great-circle arguments. The formulation is intended to complement, rather than compete with, more detailed operational weather-routing and techno-economic tools.

More concretely, we address the following questions:
\begin{enumerate}
  \item How much does enforcing strict sea-only feasibility on a realistic bathymetric grid increase trans-Arctic route length relative to a great-circle baseline?
  \item To what extent do simple depth thresholds, representing under-keel clearance for deep-draught vessels, further fragment or constrain basin-scale connectivity at \(0.5^\circ\) resolution?
  \item How do summer 2018 sea-ice conditions control the seasonal opening and closing of an ice-safe trans-Arctic corridor, and what distance penalty do they impose relative to the sea-only benchmark?
  \item How sensitive are these diagnostics to the assumed sea-ice concentration threshold used to define ice-safe waters?
  \item What is the combined impact of bathymetry and sea ice when both depth and ice-safety constraints are enforced simultaneously?
\end{enumerate}

The contributions of the paper are twofold. Methodologically, it proposes a transparent, reproducible workflow for constructing and interrogating safety-constrained routing masks on a pan-Arctic grid, using widely available bathymetric and sea-ice products and a standard graph-search algorithm. From an application perspective, it quantifies how static bathymetry and summer 2018 ice fields jointly shape the feasibility, timing and excess distance of trans-Arctic routes in a summer season, and delineates parameter ranges (season, ice threshold, depth constraint) for which basin-scale connectivity is promoted or suppressed. The framework is designed to serve as a baseline for subsequent work that will add time-dependent forcing, vessel performance models and multi-objective cost functions.
\section{Data and methods}\label{sec:methods}

\subsection{Study domain and routing grid}
\label{sec:domain_grid}

The analysis focuses on a pan-Arctic corridor that is sufficiently wide
to encompass candidate trans-Arctic connections between northern Europe
and the North Pacific, while remaining compact enough for systematic
grid-based diagnostics. We define a latitude band from
$40^{\circ}\,\mathrm{N}$ to $85^{\circ}\,\mathrm{N}$ and a longitude range
from $20^{\circ}\,\mathrm{W}$ to $180^{\circ}\,\mathrm{E}$ (hereafter
referred to as the \emph{routing domain}). This choice captures the
Norwegian and Barents Seas, the Arctic Ocean, and the Bering and
Okhotsk Seas, together with the adjacent shelves and marginal seas that
are relevant for Europe--Asia traffic, but excludes lower-latitude
subtropical waters that are not part of realistic Arctic diversions.

Within this domain we construct a regular latitude--longitude routing
grid with a resolution of $\Delta\varphi=\Delta\lambda=0.5^{\circ}$,
corresponding to a nominal horizontal spacing of approximately 55 km at Arctic mid-latitudes. Grid-cell centres
are defined on
\begin{align*}
  \varphi_i &= 40.5^{\circ}\,\mathrm{N},\, 41.0^{\circ}\,\mathrm{N},\,\dots,\,84.5^{\circ}\,\mathrm{N}, \\
  \lambda_j &= 19.5^{\circ}\,\mathrm{W},\, 19.0^{\circ}\,\mathrm{W},\,\dots,\,179.5^{\circ}\,\mathrm{E},
\end{align*}
yielding $89 \times 399$ grid points. This resolution is a compromise:
it is coarse enough to allow efficient A* routing and seasonal
sensitivity experiments over the full pan-Arctic corridor, yet fine
enough to resolve the major shelf breaks, islands and basin-scale
geometrical constraints that modulate trans-Arctic connectivity.

All routing computations operate on this coarse grid. High-resolution
datasets (bathymetry and sea-ice concentration) are first subsetted to
the routing domain and then mapped onto the $0.5^{\circ}$ grid via
bilinear interpolation in latitude--longitude space
(Sections~\ref{sec:bathy_data} and \ref{sec:ice_data}). The resulting
fields provide, for each routing cell, (i) a static ocean depth,
(ii) a static land/sea mask derived from bathymetry, and (iii) a
time-dependent sea-ice concentration. These fields are then combined
into binary feasibility masks that encode which grid cells may be used
by the pathfinding algorithm under different safety assumptions
(Sections~\ref{sec:constraints}--\ref{sec:routing_method}).

\subsection{Bathymetry dataset and land/sea mask}
\label{sec:bathy_data}

Static seafloor topography is taken from the GEBCO~2024 global grid
\citep{GEBCO2024}, subsetted to the routing domain
($40$--$85^{\circ}\,\mathrm{N}$, $20^{\circ}\,\mathrm{W}$--$180^{\circ}\,\mathrm{E}$).
The native product has a nominal resolution of $15$~arcsec; within the
subset, depth values (variable \texttt{elevation}) span from
approximately $-8700$~m (deep ocean) to $+5500$~m (continental terrain
and ice sheets).

For routing purposes we regrid GEBCO onto the $0.5^{\circ}$ routing
grid. For each routing cell centre $(\varphi_i,\lambda_j)$, depth is
obtained by bilinear interpolation in latitude and longitude, yielding a
depth field $h_{ij}$ with shape $89 \times 399$. We adopt the GEBCO
sign convention (negative values over the ocean, positive over land and
ice). A binary land/sea mask is then defined as
\begin{equation}
  M^{\mathrm{sea}}_{ij} =
    \begin{cases}
      1, & h_{ij} < 0~\mathrm{m},\\[3pt]
      0, & h_{ij} \ge 0~\mathrm{m},
    \end{cases}
\end{equation}
so that $M^{\mathrm{sea}}=1$ denotes ocean and $M^{\mathrm{sea}}=0$
denotes land or grounded ice. This mask defines the maximum set of
cells that can ever be used by a route; all subsequent depth and sea-ice
constraints operate within $M^{\mathrm{sea}}=1$.

To represent under-keel clearance requirements we further define
depth-thresholded sea masks. Given a minimum acceptable water depth
$h_{\min}$, we consider
\begin{equation}
  M^{\mathrm{depth}}_{ij}(h_{\min}) =
    \begin{cases}
      1, & M^{\mathrm{sea}}_{ij}=1 \ \text{and}\ h_{ij} \le -h_{\min},\\[3pt]
      0, & \text{otherwise},
    \end{cases}
\label{eq:mdepth}
\end{equation}
such that only cells deeper than $h_{\min}$ (in absolute value) remain
available. In this paper we mainly use $h_{\min}=20$~m as a conservative
yet permissive value for large commercial vessels, and treat stricter
depth constraints as a diagnostic in the Results.

\subsection{Sea-ice dataset and regridding}
\label{sec:ice_data}

Sea-ice concentration is taken from the CMEMS Arctic sea-ice
reanalysis \citep{CMEMS_SeaIce_Reanalysis}, which provides daily fields of
sea-ice concentration (SIC) and thickness on a polar stereographic grid
for the Arctic Ocean and surrounding seas. We focus on the June--September 2018 season as a recent summer with complete CMEMS coverage and typical low Arctic sea-ice extent, rather than an extreme minimum year, so that the diagnostics reflect present-day conditions without being tied to an exceptional season.

The SIC variable (\texttt{siconc}) is expressed as fractional sea-ice
concentration in $[0,1]$. For each daily time step $t_k$ we first
subset the CMEMS product to a geographic box slightly larger than the
routing domain and then regrid SIC onto the $0.5^{\circ}$ routing grid
via bilinear interpolation, yielding a time-dependent field
$s_{ijk} \in [0,1]$ on the same $89 \times 399$ grid as the bathymetry.
Grid points south of the CMEMS ice domain (for which \texttt{siconc} is
undefined) are assigned $s_{ijk}=0$, consistent with the product being
an Arctic sea-ice reanalysis.

The primary sea-ice control used in this study is a concentration
threshold $C_{\mathrm{thr}}$ expressed as a fraction of ice cover
(e.g.\ $C_{\mathrm{thr}} = 0.15$ for $15\%$ SIC). For a given date
$t_k$, we define an ice-feasibility mask
\begin{equation}
  M^{\mathrm{ice}}_{ij}(C_{\mathrm{thr}}, t_k) =
    \begin{cases}
      1, & M^{\mathrm{sea}}_{ij}=1 \ \text{and}\ s_{ijk} < C_{\mathrm{thr}},\\[3pt]
      0, & \text{otherwise},
    \end{cases}
\end{equation}
such that $M^{\mathrm{ice}}=1$ denotes ice-safe ocean at that date and
threshold. The reference threshold used throughout is
$C_{\mathrm{thr}} = 0.15$ (15\,\% SIC), using a concentration threshold close to the WMO ‘open water’ category (total sea-ice concentration < 1/10; \citet{WMO259}); additional thresholds between 5\,\% and
50\,\% are used for sensitivity experiments.

\subsection{Feasibility constraints: depth, ice and joint masks}
\label{sec:constraints}

All routing experiments are performed on binary feasibility masks that
encode which cells are available to the pathfinding algorithm. We
distinguish three types of masks:

\begin{enumerate}
  \item \textbf{Sea-only mask} $M^{\mathrm{sea}}$:
    this mask encodes pure geometric accessibility based on GEBCO
    bathymetry, with no additional safety constraints. It is used to
    construct a baseline ``sea-only'' route that avoids land but ignores
    both depth and sea-ice.

  \item \textbf{Depth-constrained masks}
    $M^{\mathrm{depth}}(h_{\min})$:
    these masks enforce under-keel clearance by excluding cells shallower
    than $h_{\min}$, as defined in Eq.~(2). In practice we
    consider $h_{\min}=20$~m as a realistic value for a deep-draught
    commercial vessel, and briefly explore stricter values
    ($h_{\min}=50$ and $200$~m) to assess how quickly trans-Arctic
    connectivity fragments at this resolution. To avoid artificially
    blocking port access, depth constraints may be relaxed in the
    immediate origin and destination cells while being enforced along
    the rest of the route.

  \item \textbf{Ice-constrained masks}
    $M^{\mathrm{ice}}(C_{\mathrm{thr}}, t_k)$:
    these masks encode the subset of ocean cells that are both
    geometric sea and have SIC below a chosen threshold on date $t_k$,
    as defined in Eq.~(3). They are used for seasonal and threshold
    sensitivity experiments in which only sea-ice is varied while
    bathymetry remains fixed.
\end{enumerate}

For completeness we also define a \emph{joint} depth-plus-ice mask
\begin{equation}
  M^{\mathrm{joint}}_{ij}(h_{\min}, C_{\mathrm{thr}}, t_k)
  = M^{\mathrm{depth}}_{ij}(h_{\min}) \,
    M^{\mathrm{ice}}_{ij}(C_{\mathrm{thr}}, t_k),
\end{equation}
representing the intersection of depth and ice constraints. At the
$0.5^{\circ}$ resolution used here a continuous joint-safe corridor does
not exist for $h_{\min}=20$~m and $C_{\mathrm{thr}}=15$\,\% in
September~2018; this lack of connectivity is itself one of the
diagnostics discussed in Section~\ref{sec:results_joint} and motivates
the decision to treat bathymetry and sea-ice constraints separately in
the main route analysis.

\subsection{Graph construction and routing algorithm}
\label{sec:routing_method}

Routing is performed on a regular-grid graph derived from the routing
grid. Each ocean cell $(i,j)$ with mask value $M_{ij}=1$ (for the mask
under consideration) is treated as a node. Undirected edges are added
between each node and its von Neumann and diagonal neighbours
(8-connectivity), provided that both endpoints are feasible
($M_{ij}=M_{i'j'}=1$). The edge cost $c_{(i,j)\to(i',j')}$ is taken as
the great-circle distance between the two cell centres, computed using
a spherical Earth approximation with radius $R=6371~\mathrm{km}$; this
ensures that the discrete cost field is consistent with physical
distance.

Routes between a chosen origin and destination are computed using the
A* algorithm. The heuristic $h(\cdot)$ is the great-circle distance from
the current node to the destination, divided by a constant reference
speed to preserve units; this heuristic is admissible and guarantees
that A* returns a globally distance-minimising path on the graph. For
masks that are purely geometric (sea-only) or purely ice-based, A* is
therefore equivalent to finding the shortest feasible sea-only route in
terms of travelled distance.

Throughout this paper we consider a canonical origin--destination (OD)
pair chosen to span the pan-Arctic basin without tying the analysis to a
specific port. The origin lies in the eastern North Atlantic at
$(40.5^{\circ}\,\mathrm{N},\,19.5^{\circ}\,\mathrm{W})$, while the
destination lies in the North Pacific at
$(69.5^{\circ}\,\mathrm{N},\,179.5^{\circ}\,\mathrm{E})$. Both
coordinates coincide with routing-grid cell centres and are located in
the largest contiguous sea component under the pure bathymetric mask,
so that a reference sea-only route exists. The great-circle distance
between these points is approximately $4149$~NM and serves as a lower
bound against which gridded sea-only and constrained routes are
compared.

\subsection{Diagnostic metrics and experiment design}
\label{sec:diagnostics}

For each feasibility mask and OD pair we extract the A*-derived path as
an ordered sequence of grid cells
$\{(i_n,j_n)\}_{n=0}^{N-1}$, together with their geographic
coordinates $(\varphi_{i_n}, \lambda_{j_n})$. From this discrete path we
compute the following diagnostics:

\begin{itemize}
  \item \textbf{Route distance} $D_{\mathrm{route}}$:
    the sum of great-circle distances along each edge of the path,
    reported in nautical miles (NM).

  \item \textbf{Great-circle distance} $D_{\mathrm{GC}}$:
    the great-circle distance between origin and destination, also in
    NM, representing the theoretical lower bound in the absence of
    land and safety constraints.

  \item \textbf{Route inflation} $\rho$:
    the ratio $D_{\mathrm{route}}/D_{\mathrm{GC}}$, which measures by how
    much the constrained route is elongated relative to the direct
    great-circle.

  \item \textbf{Depth statistics along the route}:
    for experiments that include bathymetric information we sample
    $h_{ij}$ along the route and report the minimum, mean and maximum
    depths. In addition, one-dimensional profiles and histograms are
    used to compare the depth distribution along the route with that of
    all sea cells in the corridor.

  \item \textbf{Sea-ice statistics and connectivity}:
    for ice-constrained experiments we compute (i) the fraction of sea
    cells that are ice-safe ($M^{\mathrm{ice}}=1$), (ii) the number of
    connected components of the ice-safe sea, and (iii) whether the
    origin and destination lie in the same ice-safe component. These
    diagnostics are computed both for single dates and for seasonal
    sequences.
\end{itemize}

Two families of experiments are performed:

\begin{enumerate}
  \item \textbf{Seasonal experiments}:
    for a fixed SIC threshold $C_{\mathrm{thr}}=0.15$ we evaluate
    ice-only routing on four representative summer dates
    (15~June, 15~July, 15~August and 15~September~2018). For each date
    we compute the fraction of ice-safe sea, the number of ice-safe
    components, and (if a route exists) the corresponding
    $D_{\mathrm{route}}$ and $\rho$.

  \item \textbf{Threshold-sensitivity experiments}:
    for a late-summer date we vary the SIC
    threshold over $\{5, 10, 15, 30, 50\}\,\%$ and repeat the
    diagnostics. A second experiment combines multiple dates
    (mid-June to end-September) and thresholds into a date--threshold
    matrix, from which we construct a heatmap of route availability and
    route inflation. Entries where no continuous ice-safe corridor
    exists are marked as missing, emphasising the discrete ``opening''
    of the trans-Arctic connection in late summer.
\end{enumerate}

Joint depth-plus-ice masks are analysed in a more qualitative way: for
selected combinations of $h_{\min}$ and $C_{\mathrm{thr}}$ we visualise
the joint-safe sea fraction and its connected components, and record
whether a continuous trans-Arctic corridor exists. At the
$0.5^{\circ}$ resolution considered here, no joint-safe path is found
for $h_{\min}=20$~m and $C_{\mathrm{thr}}=15$\,\% in September~2018,
highlighting the strong fragmentation induced by combining depth and
ice constraints and motivating the separation of bathymetric and
sea-ice effects in the present paper.

\subsection*{Notation for masks and thresholds}

For convenience, Table~\ref{tab:notation_masks} summarises the main
symbols used for feasibility masks and thresholds in this paper.

\begin{table}[htbp]
  \centering
  \caption{Notation for masks, fields and thresholds used in the routing framework.}
  \label{tab:notation_masks}
  \begin{tabular}{llp{7cm}}
    \hline
    Symbol & Type & Meaning \\
    \hline
    $h_{ij}$ &
    scalar field &
    Water depth at routing-grid cell $(i,j)$ from GEBCO~2024; negative over ocean, positive over land/ice. \\[3pt]

    $M^{\mathrm{sea}}_{ij}$ &
    binary mask &
    Static land/sea mask; $M^{\mathrm{sea}}_{ij}=1$ for ocean cells ($h_{ij}<0$), $0$ otherwise. \\[3pt]

    $h_{\min}$ &
    parameter &
    Minimum acceptable water depth (under-keel clearance), e.g.\ $20$~m, $50$~m or $200$~m. \\[3pt]

    $M^{\mathrm{depth}}_{ij}(h_{\min})$ &
    binary mask &
    Depth-safe sea mask; $M^{\mathrm{depth}}_{ij}=1$ if $M^{\mathrm{sea}}_{ij}=1$ and $h_{ij}\le -h_{\min}$. \\[3pt]

    $s_{ijk}$ &
    scalar field &
    Sea-ice concentration (SIC) at cell $(i,j)$ and time index $k$ regridded from CMEMS, in $[0,1]$. \\[3pt]

    $C_{\mathrm{thr}}$ &
    parameter &
    SIC threshold for ice safety, expressed as fraction (e.g.\ $0.15$ for $15\%$). \\[3pt]

    $M^{\mathrm{ice}}_{ij}(C_{\mathrm{thr}}, t_k)$ &
    binary mask &
    Ice-safe sea mask at date $t_k$; $M^{\mathrm{ice}}_{ij}=1$ if $M^{\mathrm{sea}}_{ij}=1$ and $s_{ijk}<C_{\mathrm{thr}}$. \\[3pt]

    $M^{\mathrm{joint}}_{ij}(h_{\min}, C_{\mathrm{thr}}, t_k)$ &
    binary mask &
    Joint depth-plus-ice mask; product of $M^{\mathrm{depth}}_{ij}(h_{\min})$ and $M^{\mathrm{ice}}_{ij}(C_{\mathrm{thr}}, t_k)$. \\[3pt]

    $D_{\mathrm{GC}}$ &
    scalar &
    Great-circle distance between origin and destination (NM). \\[3pt]

    $D_{\mathrm{route}}$ &
    scalar &
    Distance of the A*-derived route on the grid (NM). \\[3pt]

    $\rho$ &
    scalar &
    Route inflation factor, $\rho = D_{\mathrm{route}} / D_{\mathrm{GC}}$. \\
    \hline
  \end{tabular}
\end{table}

\newpage
\section{Results and Discussion}\label{sec:results_and_discussion}

\subsection{Bathymetric domain and baseline sea-only routing}
\label{sec:results_bathy_baseline}

\begin{figure}[htbp]
  \centering
  \includegraphics[width=\linewidth]{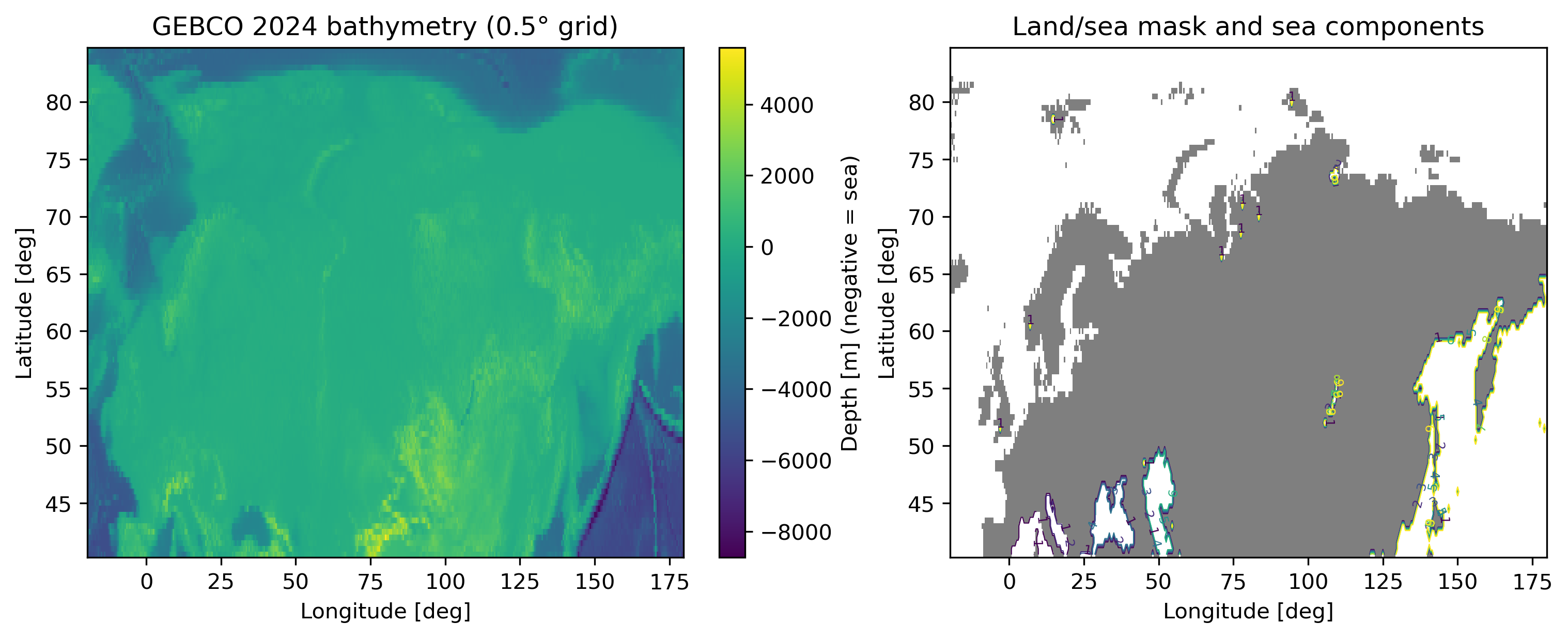}
  \caption{%
  (a) GEBCO~2024 bathymetry coarsened to a $0.5^{\circ}\times0.5^{\circ}$ routing
  grid over the Arctic corridor (40--85$^{\circ}$N, 20$^{\circ}$W–180$^{\circ}$E).
  (b) Derived land/sea mask and connected sea components; the largest sea
  component links the North Atlantic and North Pacific, while smaller ones are
  enclosed shelf seas and fjords.%
  }
  \label{fig:bathy_domain_components}
\end{figure}

We first define a static bathymetric routing grid for the trans-Arctic corridor.
Bathymetry is taken from the GEBCO\_2024 global grid and subset to the region
$40$--$85^{\circ}$N and $20^{\circ}$W–$180^{\circ}$E, which captures the North
Atlantic, Barents and Kara seas, the Eurasian Arctic shelf and the Bering and
North Pacific gateway. The native GEBCO field (15~arcsec) is coarsened to a
regular $0.5^{\circ}\times0.5^{\circ}$ latitude--longitude grid yielding a routing grid of $89\times399$ nodes
(Fig.~\ref{fig:bathy_domain_components}a). Within this domain, depths on the
routing grid range from approximately $-8700$~m in the deep basins to $+5600$~m
over land.

A binary land/sea mask is constructed by classifying grid cells with
GEBCO elevation $\leq 0$~m as sea and $>0$~m as land
(Fig.~\ref{fig:bathy_domain_components}b). Connected-component labelling on the
sea mask shows that the largest sea component forms a continuous oceanic
corridor from the North Atlantic to the North Pacific across the Arctic, while
smaller components correspond to enclosed shelf seas and fjords. This mask
defines the static sea-only feasibility constraint used throughout the
remainder of the study.

\begin{figure}[htbp]
  \centering
  \includegraphics[width=0.75\linewidth]{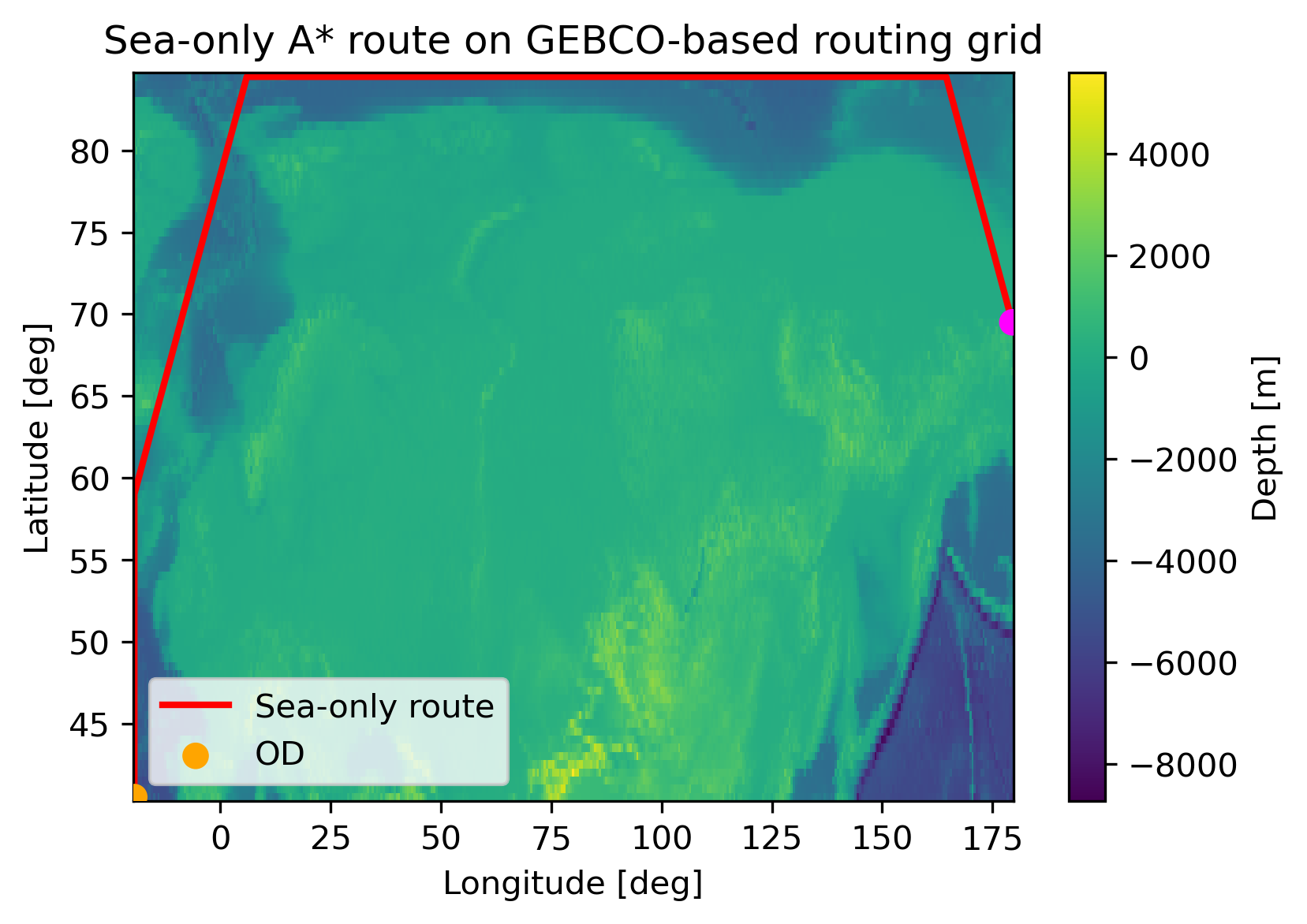}
  \caption{%
  Sea-only A* route (red) between the canonical origin--destination pair
  overlaid on the GEBCO-based $0.5^{\circ}$ bathymetry. The route remains within
  the static sea mask and follows deep-water pathways along the Arctic margin.%
  }
  \label{fig:seaonly_route}
\end{figure}

To obtain a canonical baseline route, we consider an idealised
trans-Arctic origin–destination (OD) pair between a North Atlantic waypoint at
$(40.5^{\circ}\text{N},\,19.5^{\circ}\text{W})$ and a North Pacific waypoint at
$(69.5^{\circ}\text{N},\,179.5^{\circ}\text{E})$, corresponding to the
south-western and north-eastern corners of the routing domain. The great-circle
distance between these two waypoints is $4149.6$~NM. Routing on the $0.5^{\circ}$
grid is performed using an A* search with edge costs equal to the great-circle
distance between neighbouring grid nodes, restricted to cells in the static
sea mask. The resulting sea-only route is shown in
Fig.~\ref{fig:seaonly_route} and runs northwards out of the North Atlantic,
skirts the Eurasian Arctic shelf break and exits into the North Pacific.

\begin{figure}[htbp]
  \centering
  \includegraphics[width=\linewidth]{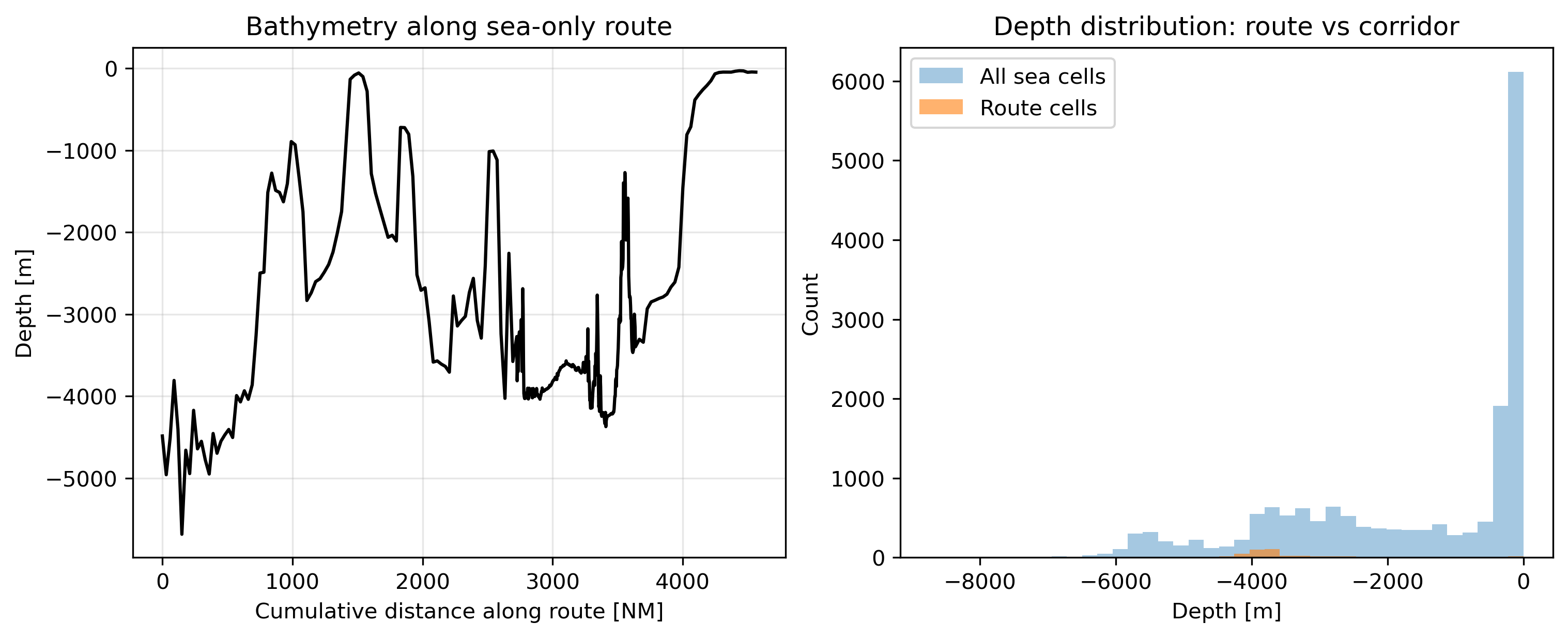}
  \caption{%
  (a) Bathymetry along the baseline sea-only route as a function of cumulative
  distance. (b) Depth histograms for all sea cells in the corridor (blue) and
  for routed cells (orange); the route samples predominantly deep water and
  only briefly crosses shallow shelves.%
  }
  \label{fig:depth_profile}
\end{figure}

The sea-only route distance is $4561.1$~NM, i.e.\ approximately $10\%$ longer
than the unconstrained great-circle distance. At a design speed of 14~kn this
corresponds to a travel time of 13.6~days, compared to 12.4~days for the
great-circle, implying a penalty of about 1.2~days purely due to enforcing
sea-only feasibility on a coarse grid. Bathymetry sampled along the route
(Fig.~\ref{fig:depth_profile}) indicates that the path remains predominantly in
deep water: depths along the route have a mean of $\approx -3300$~m, with a
minimum of $\approx -5700$~m and a shallowest segment of $\approx -30$~m. The
associated depth histogram shows that the route samples the deeper part of the
corridor-wide depth distribution, with only a few segments approaching the
continental shelf. This confirms that the purely geometric sea-only constraint
already tends to select deep-water pathways, even before any explicit
bathymetric safety threshold is applied.

\subsection{Bathymetric safety thresholds and routing connectivity}
\label{sec:results_depth}

Having established a purely geometric sea-only route on the GEBCO-based grid,
we next examine how explicit bathymetric safety thresholds affect network
connectivity and route geometry. For a given minimum safe depth
\(h_\mathrm{min}\), the depth-safe sea mask \(M^\mathrm{depth}_{ij}(h_\mathrm{min})\) is
defined in Eq.~\eqref{eq:mdepth} as taking the value 1 where cell \((i,j)\) is
ocean and \(h_{ij} \le -h_\mathrm{min}\), and 0 otherwise; that is, only grid cells
deeper than \(h_\mathrm{min}\) are considered navigable.

In practice we apply this threshold to all interior cells of a candidate route,
while allowing the origin and destination cells themselves to be shallower
(``endpoints relaxed''), mimicking the fact that vessels typically enter or
leave deep-water routes through local port approaches.

\begin{figure}[htbp]
  \centering
  \includegraphics[width=\linewidth]{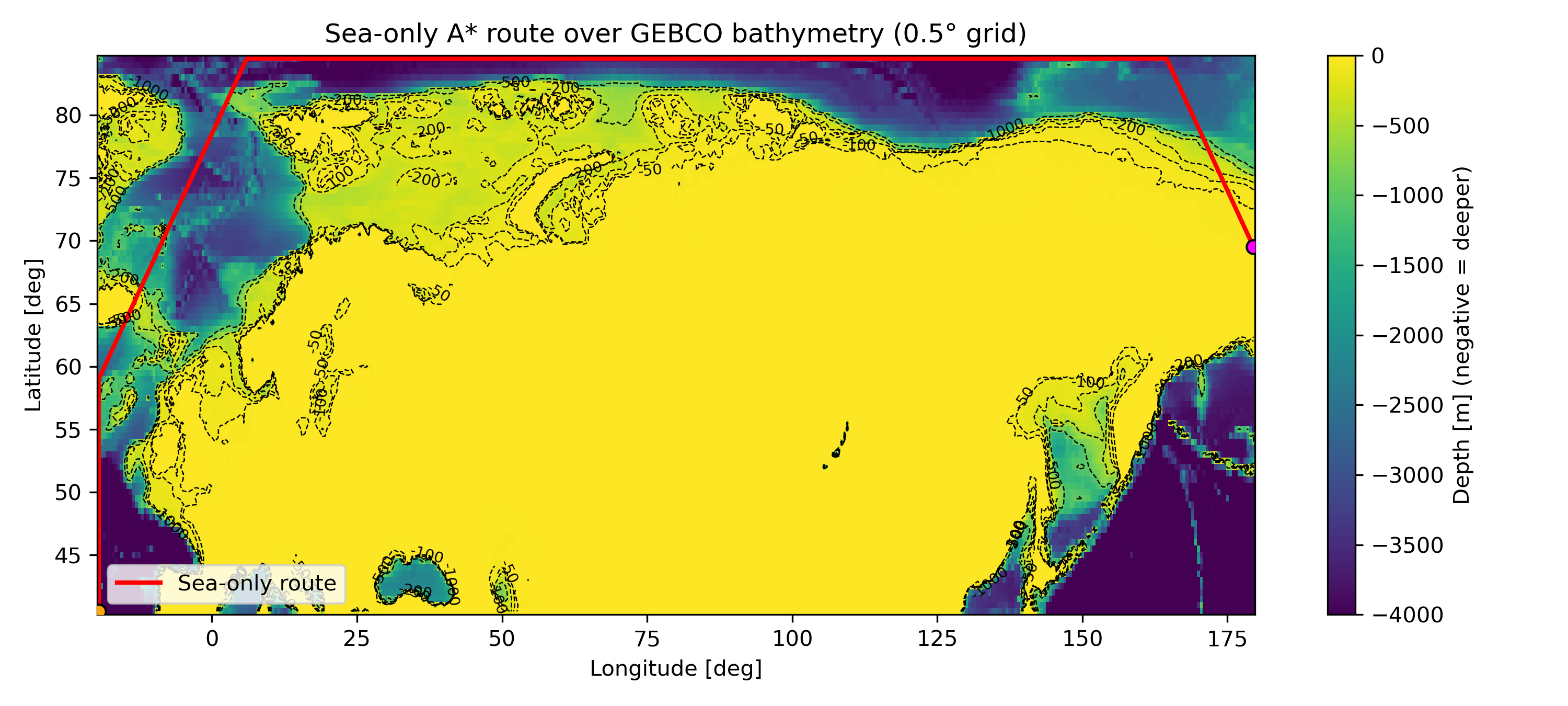}
  \caption{%
  Sea-only A* route (red) over GEBCO~2024 bathymetry on the $0.5^{\circ}$ grid,
  with selected isobaths. The path tracks the Arctic shelf break and remains
  mostly in deep water even without explicit depth constraints.%
  }
  \label{fig:depth_path_map}
\end{figure}

On the $0.5^{\circ}$ routing grid, the static sea mask occupies
approximately $48\%$ of all cells. Imposing a relatively mild depth constraint
of $h_{\min}=20$~m retains $94\%$ of this sea area as depth-safe
($\approx 45\%$ of the whole grid) and leaves the canonical trans-Arctic
origin and destination in the same connected component. Running A* on this
depth-safe mask yields a route that is effectively identical to the baseline
sea-only path: the distance remains $4561$~NM, corresponding to a $10\%$
inflation over the great-circle distance, and depths sampled along the route
remain almost entirely in deep water
(Fig.~\ref{fig:depth_path_map}, cf.~Fig.~\ref{fig:depth_profile}). This confirms that, at the scale of our
corridor, the unconstrained sea-only route already follows the deep-ocean
backbone and does not exploit marginally shallow shortcuts.

\begin{figure}[htbp]
  \centering
  \includegraphics[width=0.9\linewidth]{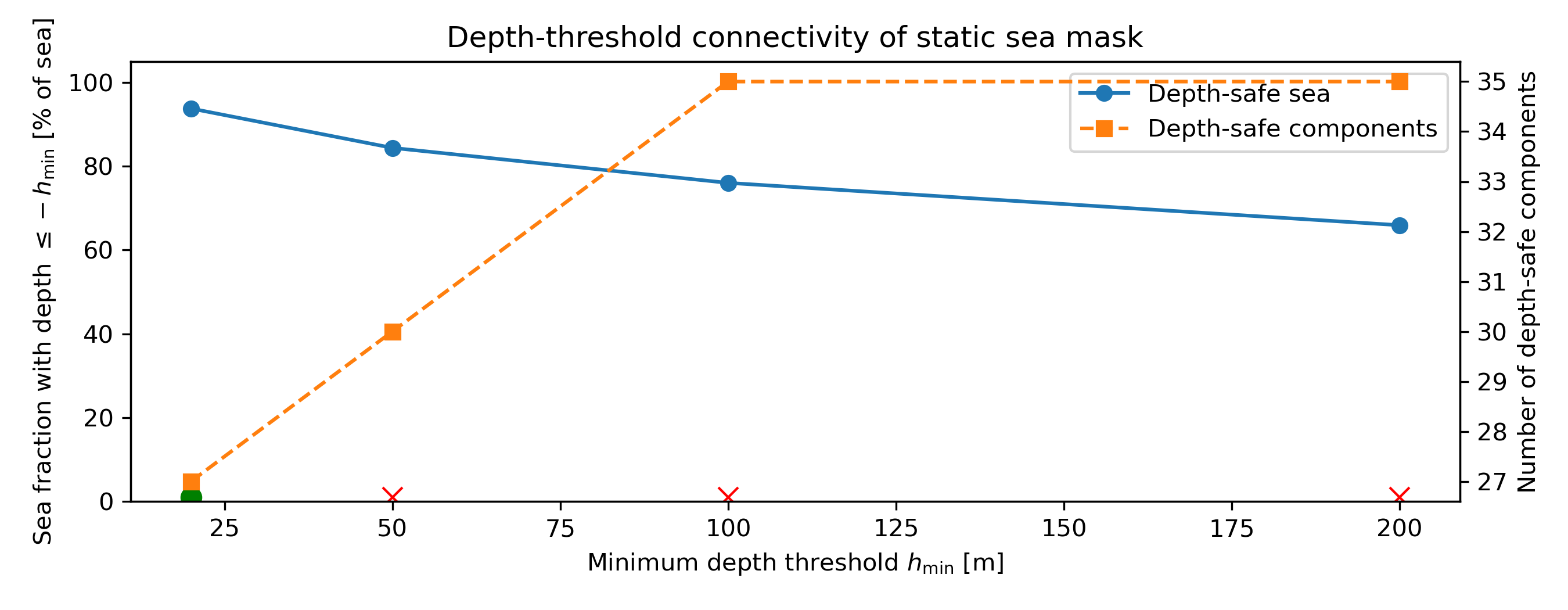}
  \caption{%
  Effect of minimum depth $h_{\min}$ on the fraction of depth-safe sea area
  (blue, left axis) and the number of depth-safe components (orange, right
  axis). Markers indicate whether the canonical origin and destination lie in
  the same component (route possible) or not.%
  }
  \label{fig:depth_connectivity}
\end{figure}

As the minimum safe depth is tightened, the navigable network gradually
fragments. For $h_{\min}=50$~m, only about $85\%$ of the original sea area
remains depth-safe; by $h_{\min}=200$~m this drops to roughly two-thirds of the
sea mask. The number of depth-safe connected components increases from
$30$ to more than $35$ as the threshold is raised, and beyond
$h_{\min}\approx 50$~m the canonical origin and destination lie in different
depth-safe components (Fig.~\ref{fig:depth_connectivity}b). In other words, at
this resolution there is no continuous path between the Atlantic and Pacific
gateways that is everywhere deeper than 50--200~m.

The spatial pattern of these constraints is illustrated in
Fig.~\ref{fig:depth_path_map}a, where the sea-only route is overlaid on
bathymetry with selected isobaths. Narrow shallow features along the shelf
break and in straits appear as ``choke points'' that disconnect the deep-water
network once a stringent depth threshold is enforced. From an operational
perspective, vessels with realistic draughts could still transit these regions,
but at the $0.5^{\circ}$ grid spacing they collapse into single grid cells that
are either fully safe or fully blocked. This highlights an important trade-off:
coarser grids are computationally attractive for large-scale scenario analysis,
but they can overstate the impact of bathymetric constraints in narrow
passages.

A complementary diagnostic is provided by the depth distribution in the
corridor versus along the route
(Fig.~\ref{fig:depth_profile}). While the overall sea area spans depths from
approximately $-100$ to $-6000$~m, the A* route samples predominantly the
deeper part of this distribution, with a long tail towards abyssal depths and
only a small fraction of cells shallower than 100~m. Together with the
connectivity analysis, this suggests that in the present configuration
bathymetry is \emph{not} the primary limiting factor for a trans-Arctic
corridor at the coarse planning scale; rather, the static sea-only feasibility
and the seasonal sea-ice field (addressed in the following sections) are the
dominant controls on route availability.

\subsection{Seasonal opening of an ice-safe trans-Arctic corridor}
\label{sec:results_seasonal_ice}

We now investigate how seasonal sea-ice concentration modulates the feasibility
and geometry of the trans-Arctic corridor on the static bathymetric grid.
Sea-ice concentration (SIC) is taken from the CMEMS Arctic sea-ice reanalysis
for June--September 2018 and regridded onto the $0.5^{\circ}$ routing grid by
bilinear interpolation. For each target date we extract the nearest model time,
apply a fixed ice-safety criterion of $\mathrm{SIC}<15\%$, and intersect this
ice-safe mask with the static sea mask. The resulting ice-safe sea area and
associated connectivity are then evaluated for the canonical origin--destination
pair.

\begin{figure}[htbp]
  \centering
  \includegraphics[width=0.9\linewidth]{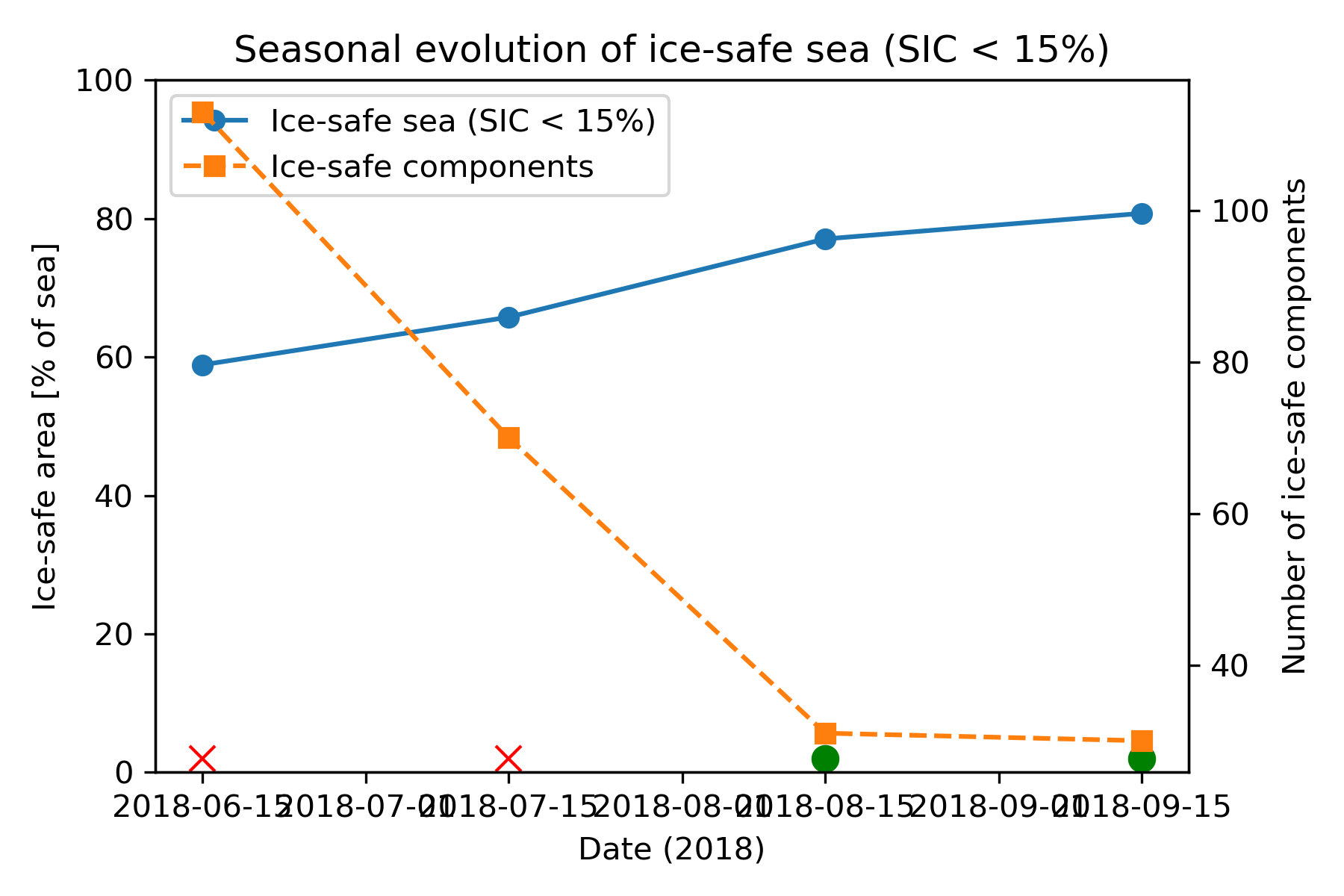}
  \caption{%
  Seasonal evolution of the ice-safe sea fraction (blue, left axis) and number
  of ice-safe components (orange, right axis) for $\mathrm{SIC}<15\%$. Markers
  show dates when the canonical origin and destination lie in the same
  ice-safe component (route exists) or not.%
  }
  \label{fig:seasonal_ice_area}
\end{figure}

Figure~\ref{fig:seasonal_ice_area} shows the evolution of the ice-safe sea
fraction between mid-June and mid-September 2018, expressed as a percentage of
the static sea mask. In mid-June (15 June), only $58.6\%$ of the sea area is
ice-safe under the $\mathrm{SIC}<15\%$ criterion and the ice-safe domain is
fragmented into 96 disconnected components. By mid-July the ice-safe fraction
has increased to $65.5\%$ and the number of components has decreased to 66, but
no continuous ice-safe path exists between the Atlantic and Pacific gateways.
A qualitatively different regime emerges in late summer: by 15 August,
$77.0\%$ of the sea area is ice-safe and the number of components has dropped
to 33, at which point the canonical origin and destination first fall within
the same ice-safe component. By 15 September the ice-safe fraction reaches
$80.7\%$ of the static sea mask, still partitioned into 33 components, but with
a robust large-scale path spanning the Arctic.
These seasonal changes in ice-safe area, connectivity and route availability
are summarised in Table~\ref{tab:seasonal_ice_metrics}.

\begin{table}[htbp]
  \centering
  \caption{Seasonal evolution of ice-safe sea area and route availability for the
  canonical trans-Arctic origin--destination pair under the $\mathrm{SIC}<15\%$
  criterion. Ice-safe fractions are expressed relative to the static sea mask.
  ``Route'' indicates whether a continuous ice-safe path exists.}
  \label{tab:seasonal_ice_metrics}
  \resizebox{\linewidth}{!}{%
    \begin{tabular}{lccccc}
      \hline
      Date (2018) &
      Ice-safe sea [\% of sea] &
      Ice-safe components &
      Route &
      $D_{\mathrm{route}}$ [NM] &
      $\rho = D_{\mathrm{route}}/D_{\mathrm{GC}}$ \\
      \hline
      15 June      & 58.6 & 96 & No  & --     & --    \\
      15 July      & 65.5 & 66 & No  & --     & --    \\
      15 August    & 77.0 & 33 & Yes & 5206.0 & 1.255 \\
      15 September & 80.7 & 33 & Yes & 5019.9 & 1.210 \\
      \hline
    \end{tabular}%
  }
\end{table}

\begin{figure}[htbp]
  \centering
  \includegraphics[width=0.9\linewidth]{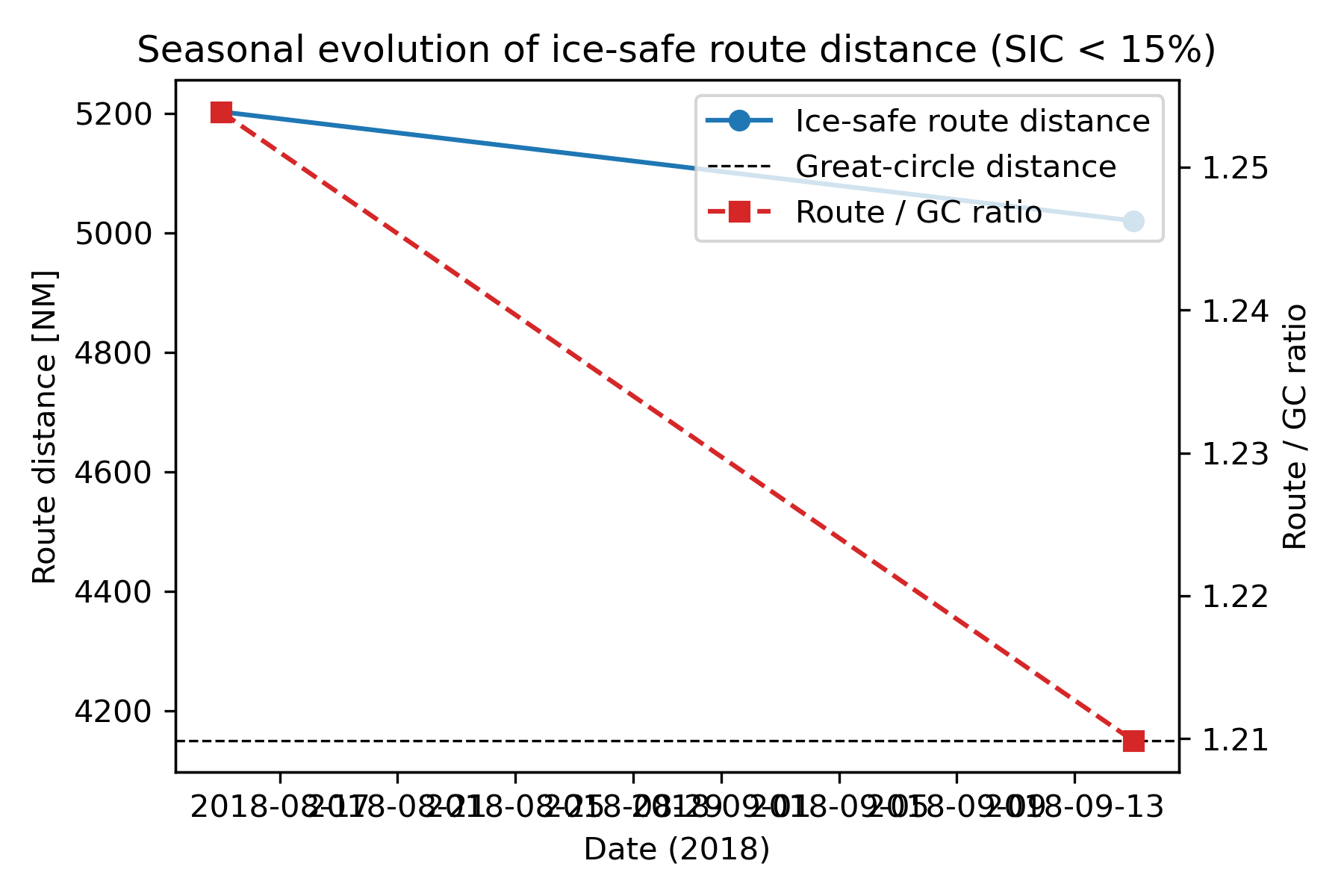}
  \caption{%
  Seasonal evolution of the shortest ice-safe route distance (circles, left
  axis) and route-to-great-circle ratio (red dashed line, right axis) for
  $\mathrm{SIC}<15\%$. No route exists in June and mid-July; from mid-August
  the corridor opens and gradually shortens as ice retreats.%
  }
  \label{fig:seasonal_route_distance}
\end{figure}

The impact of this seasonal evolution on routing is summarised in
Fig.~\ref{fig:seasonal_route_distance}. For the early-summer dates (15 June and
15 July) no ice-safe route exists under the $\mathrm{SIC}<15\%$ threshold, i.e.\
the trans-Arctic corridor is effectively \emph{closed} for the chosen
origin--destination pair despite the existence of a deep-water backbone. From
15 August onwards, an ice-safe sea-only route becomes available. On 15 August
its length is $5206$~NM, corresponding to a route-to-great-circle ratio of
$1.255$; by 15 September the distance has shortened to $5019.9$~NM and the
ratio to $1.210$. For comparison, the static bathymetry-based sea-only route
(ignoring ice) is $4561.1$~NM with a ratio of $1.099$ relative to the
$4149.6$~NM great-circle distance. Thus, even once the corridor is open in
late summer, the ice-safe route remains approximately $10\%$ longer than the
static sea-only path and about $20$--$25\%$ longer than the great-circle.

\begin{figure}[htbp]
  \centering
  \includegraphics[width=\linewidth]{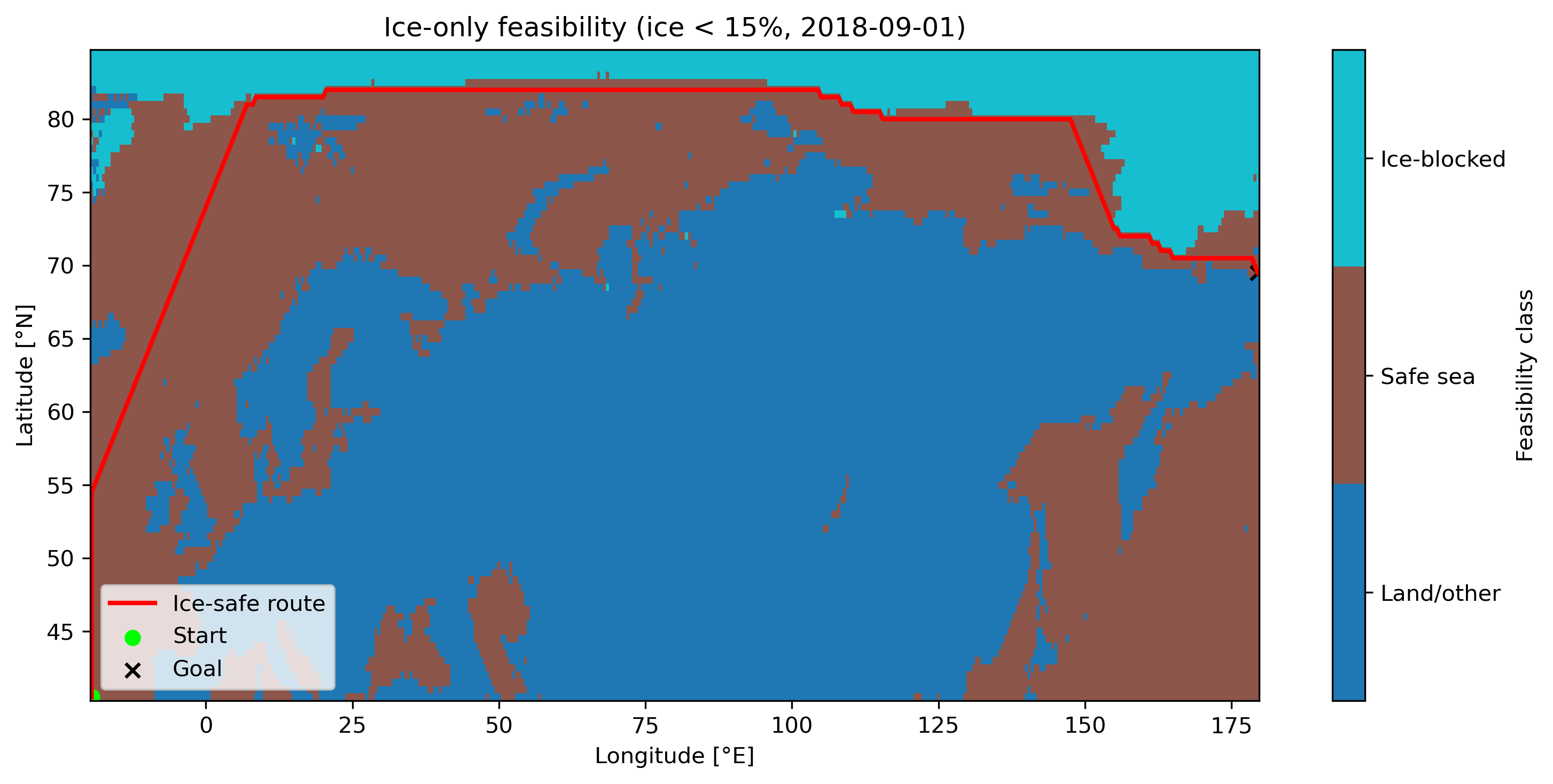}
\caption{%
Ice-only feasibility on 15~September~2018 for the canonical origin--destination
pair under the $\mathrm{SIC}<15\%$ criterion. Brown shading shows ice-safe sea,
cyan shows sea blocked by ice, and blue indicates land or non-sea. The red line
is the shortest ice-safe route, with origin and destination marked.%
}
  \label{fig:ice_feas_map}
\end{figure}

An illustrative late-summer ice-feasibility map and corresponding shortest
route are shown in Fig.~\ref{fig:ice_feas_map} for
15~September~2018. In mid-August, residual ice in the central Arctic and
along parts of the Siberian margin forces the route to skirt the ice edge,
yielding a strongly curved path that stays close to the deep
Norwegian--Greenland Sea and the Pacific side of the Arctic basin. By
mid-September, the ice extent has retreated further towards the central
Arctic and the marginal seas are largely accessible, allowing the route to
cut more directly across the basin and shorten by almost 200~NM. Despite
this shortening, the corridor remains substantially longer than both the
bathymetry-only path and the unconstrained great-circle.

Taken together, these results highlight that for a given bathymetric grid the
presence of sea ice imposes a \emph{binary} control on trans-Arctic feasibility
at seasonal timescales. For early-summer conditions in 2018 the chosen
origin--destination pair cannot be connected by any continuous ice-safe path
under a conservative $\mathrm{SIC}<15\%$ threshold, even though a deep-water
backbone exists. Once the corridor opens in late summer, sea-ice still adds a
second-order but non-negligible distance penalty on top of the static sea-only
inflation. This suggests that, from the perspective of large-scale voyage
planning, the timing of the seasonal ice retreat is a more critical driver of
trans-Arctic feasibility than moderate variations in the exact ice-safety
threshold, an aspect further explored in the threshold-sensitivity analysis in
Section~\ref{sec:results_ice_threshold}.

\subsection{Sensitivity to the ice-safety threshold}
\label{sec:results_ice_threshold}

The previous section adopted a conservative but conventional ice-safety
criterion of $\mathrm{SIC}<15\%$ to define navigable sea. Here we examine how
sensitive the large-scale routing results are to this choice by varying the
threshold over a broad range of values and repeating the analysis.

\begin{figure}[htbp]
  \centering
  \includegraphics[width=0.9\linewidth]{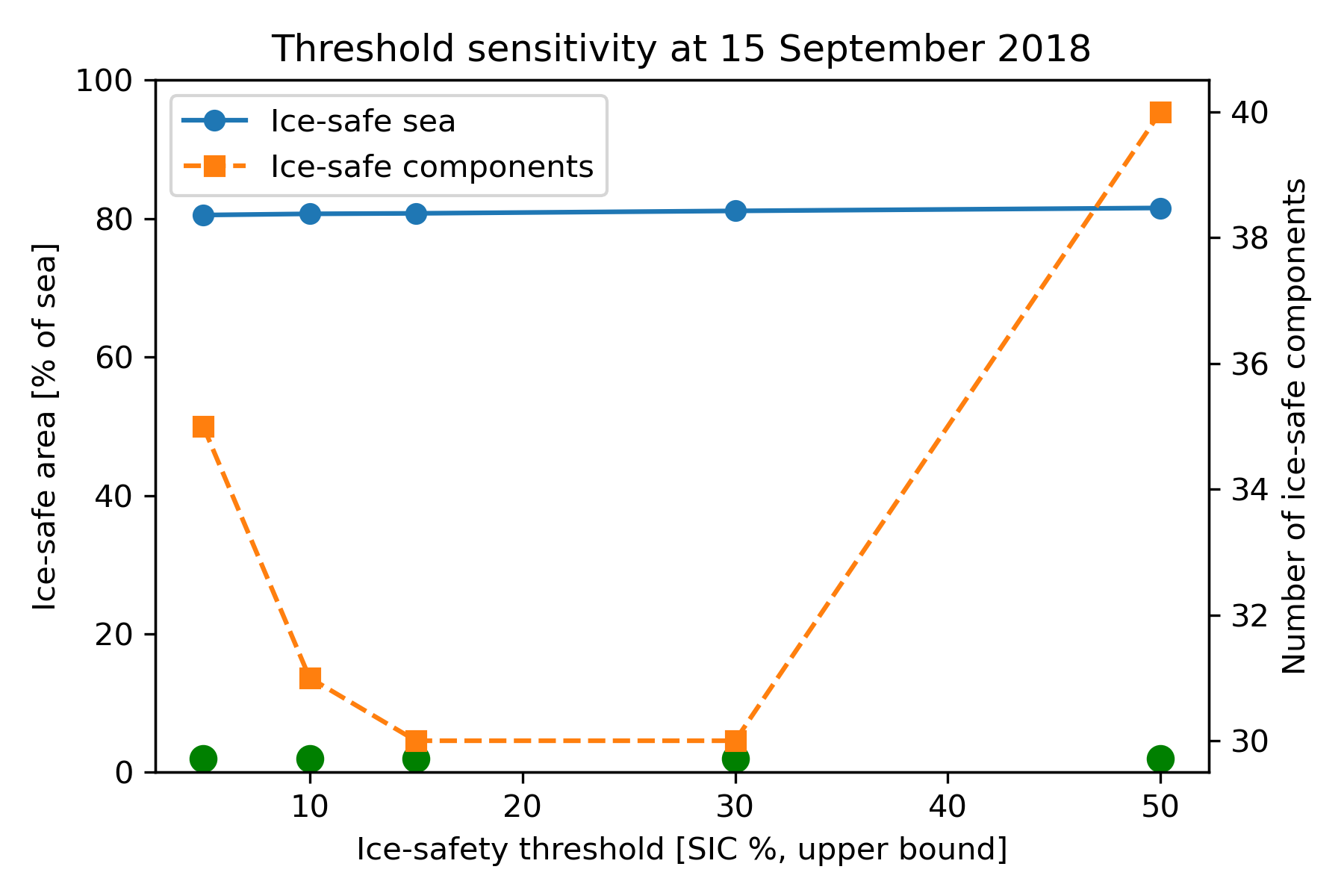}
  \caption{%
  Sensitivity of ice-safe routing to the SIC threshold on 15~September~2018:
  fraction of ice-safe sea (blue, left axis), number of ice-safe components
  (orange, right axis), and OD connectivity (markers).%
  }
  \label{fig:ice_threshold_curves}
\end{figure}

We first fix the date to 15~September~2018, when the canonical
origin--destination pair is connected by an ice-safe route for the reference
threshold, and vary the ice-safety cut-off across $\mathrm{SIC}<5\%$, $10\%$,
$15\%$, $30\%$ and $50\%$. For each threshold we recompute the fraction of
static sea area that is ice-safe, the number of disconnected ice-safe
components, and, where possible, the shortest ice-safe A* route between the
Atlantic and Pacific gateways. As shown in
Fig.~\ref{fig:ice_threshold_curves}, the ice-safe area fraction increases only
modestly across this range, from approximately $80.4\%$ of the sea mask at
$\mathrm{SIC}<5\%$ to $81.5\%$ at $\mathrm{SIC}<50\%$. The number of
ice-safe components fluctuates between roughly 33 and 44, reflecting local
changes in the connectivity of marginal-ice zones, but the canonical origin
and destination remain in the same ice-safe component for all thresholds
considered.

The resulting route distances are remarkably insensitive to the choice of
threshold. For 15~September~2018, the shortest ice-safe sea-only path ranges
from $5024.5$~NM at $\mathrm{SIC}<5\%$ to $5018.6$~NM at
$\mathrm{SIC}<50\%$, corresponding to route-to-great-circle ratios between
$1.211$ and $1.209$. This spread of less than 6~NM and $<0.2\%$ in ratio is
negligible compared to the $\approx 10\%$ inflation relative to the
bathymetry-only sea route and the $\approx 20\%$ inflation relative to the
unconstrained great-circle distance. In other words, once the large-scale
trans-Arctic corridor is open, moderate tightening or relaxation of the
ice-safety threshold within the range 5--50\% has only a second-order effect on
route length.

\begin{figure}[htbp]
  \centering
  \includegraphics[width=\linewidth]{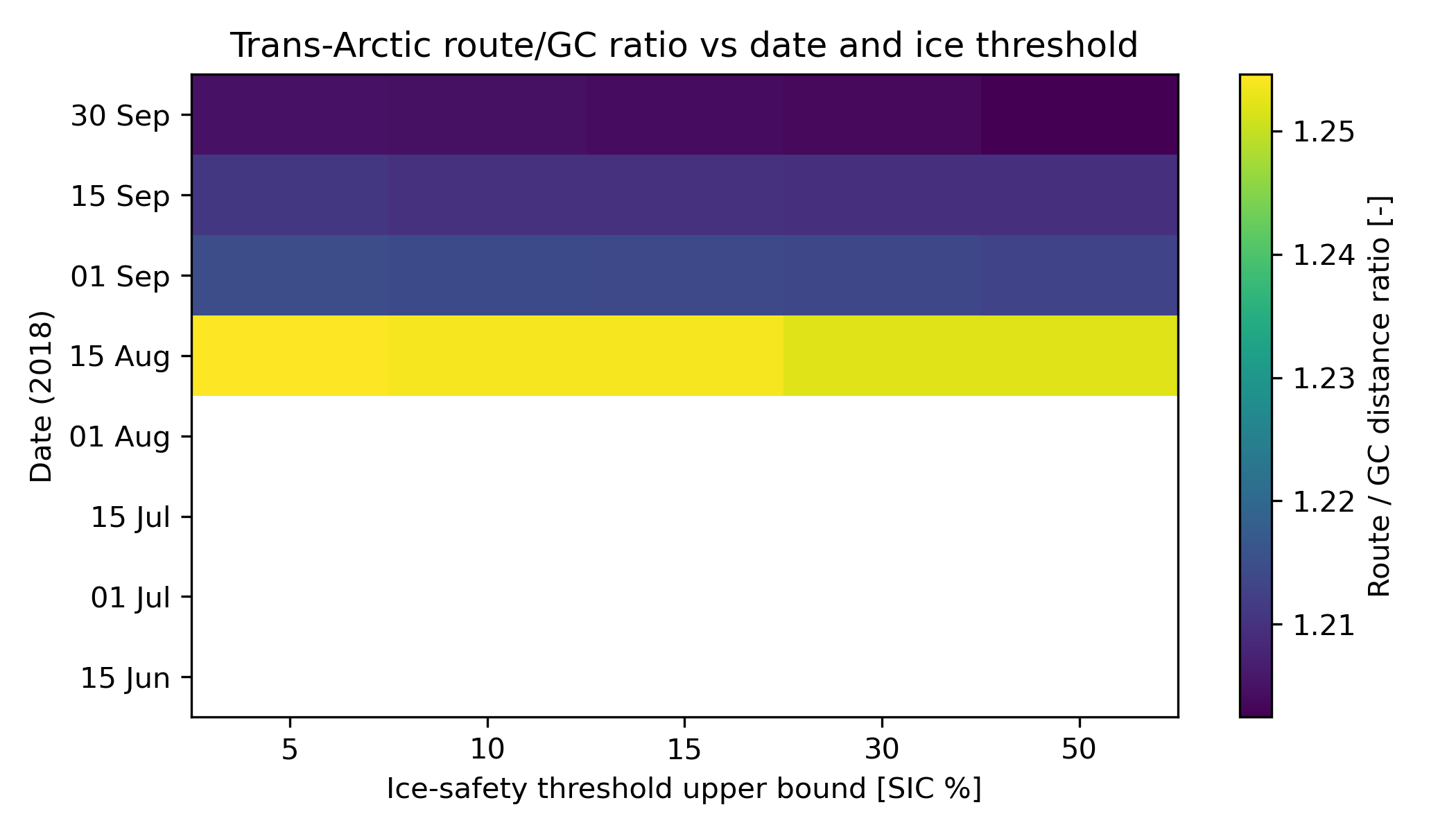}
  \caption{%
  Route-to-great-circle distance ratio for the canonical origin--destination
  pair as a function of date and SIC threshold. Colours show ratios for the
  shortest ice-safe routes; white crosses indicate combinations with no
  continuous ice-safe path.%
  }
  \label{fig:heatmap_date_threshold_ratio}
\end{figure}

To place this in a broader seasonal context, we extend the analysis to a grid
of dates and thresholds. For a set of representative dates between
15~June~2018 and 30~September~2018, and for each of the five thresholds, we
recompute the ice-safe mask and attempt to route the canonical
origin--destination pair. The resulting route-to-great-circle ratios are
summarised in Fig.~\ref{fig:heatmap_date_threshold_ratio}. Early in the
season (mid-June and early July), no continuous ice-safe path exists for any of
the thresholds in the 5--50\% range: the canonical gateways lie either outside
the ice-safe domain or in different disconnected components. The corridor thus
remains effectively closed, independently of the exact choice of
$\mathrm{SIC}$ cut-off.

A qualitatively different regime emerges once the seasonal ice cover retreats
sufficiently. From mid-August onwards, an ice-safe trans-Arctic route exists
for all tested thresholds, with route-to-great-circle ratios between
$\approx 1.25$ and $\approx 1.20$. The dominant gradient in the heatmap is
along the \emph{time} axis: ratios decrease from about $1.25$ in mid-August to
about $1.20$ by late September as the corridor widens and routes can cut more
directly across the basin. In contrast, variation along the threshold axis at a
given date is weak, typically at the level of a few tenths of a percent in
route length.

These results reinforce the interpretation that, at the scale of the present
corridor and grid, the seasonal timing of the ice retreat exerts a
first-order control on route feasibility and length, while the precise value
of the ice-safety threshold within a reasonable range primarily influences
local connectivity and the detailed shape of marginal-ice detours. For the
purposes of large-scale historical route diagnostics, a single reference
threshold such as $\mathrm{SIC}<15\%$ therefore appears adequate, provided
that the associated limitations for narrow passages and coastal approaches are
made explicit.

\subsubsection{Joint bathymetric and ice constraints}
\label{sec:results_joint}

The previous subsections considered bathymetric and ice constraints separately.
In a final diagnostic step we combined both into a joint feasibility mask,
requiring each routing cell to be (i) ocean, (ii) deeper than a minimum
under-keel clearance threshold $h_{\min}$, and (iii) ice-safe in the sense that
sea-ice concentration (SIC) remains below a navigational limit
$c_{\mathrm{ice}}^{\max}$.  Here we adopt $h_{\min}=20$~m and
$c_{\mathrm{ice}}^{\max}=15\%$ as representative values for unrestricted
open-water vessels and commonly used SIC thresholds in Arctic route screening.

For a late-summer snapshot (15~September~2018) the routing grid sea fraction
can be decomposed as follows.  Using only bathymetry, $93.8\%$ of the sea cells
are deeper than $20$~m, confirming that depth alone is not a primary limiting
factor at the $0.5^{\circ}$ corridor scale.  Using only ice, $80.8\%$ of sea
cells satisfy SIC~$<15\%$ on that date.  When both constraints are applied
simultaneously, the fraction of \emph{joint-safe} sea decreases to $74.6\%$ of
the sea mask.  While this reduction in area is modest, its spatial
organisation changes markedly: the joint-safe mask is split into
$35$ disconnected components on the $0.5^{\circ}$ grid.

\begin{figure}[t]
  \centering
  \includegraphics[width=\textwidth]{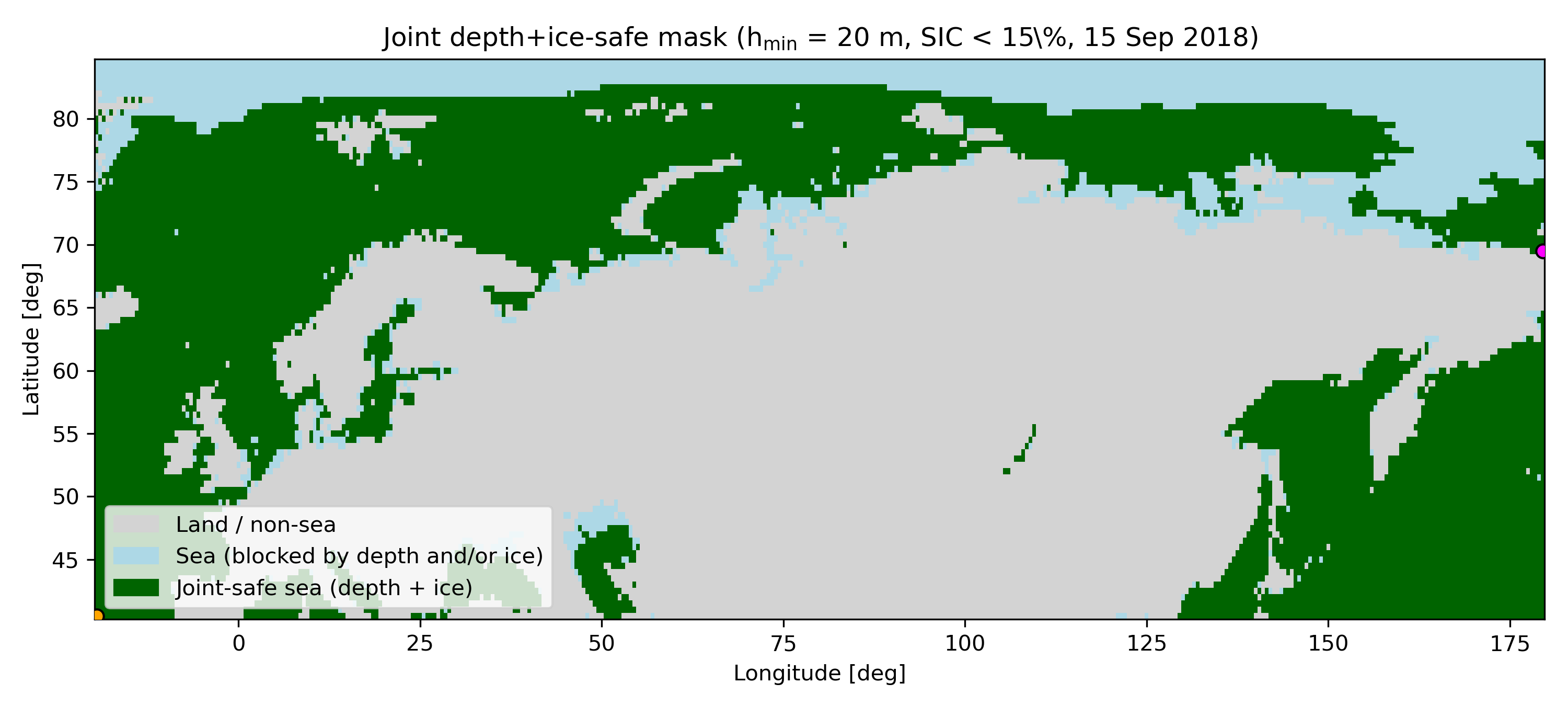}
  \caption{%
  Joint depth--ice feasibility mask ($h \ge 20$~m, SIC~$<15\%$) on the
  $0.5^{\circ}$ routing grid for 15~September~2018. Land, blocked sea and
  joint-safe sea are shown; the canonical origin and destination lie in
  different joint-safe components, so no continuous depth- and ice-safe route
  exists.%
  }
  \label{fig:joint-depth-ice-mask}
\end{figure}

Figure~\ref{fig:joint-depth-ice-mask} shows the resulting classification of the
Arctic corridor into land, sea blocked by either depth or ice, and joint-safe
sea.  The canonical trans-Arctic OD endpoints lie in different joint-safe
components (labels~1 and~20), so that no continuous route exists that
simultaneously respects $h_{\min}=20$~m and SIC~$<15\%$ along the entire path.
This is in stark contrast to the sea-only and ice-only experiments, where
late-summer Atlantic--Pacific connections could still be traced across the
same grid Table~\ref{tab:route_scenarios}. 

These results highlight that, even in a relatively favourable ice year such as
2018, the combination of shallow shelves and residual marginal ice is capable
of severing trans-Arctic connectivity at coarse resolution for realistic
under-keel and SIC thresholds.  In other words, large-scale
bathymetry-conditioned, ice-only route diagnostics (Sections~\ref{sec:results_depth}
and~\ref{sec:results_seasonal_ice}) are optimistic upper bounds on what can be
achieved when both constraints are enforced simultaneously.  From an
operational perspective this supports a two-step interpretation: (i) the
present paper’s sea-only and ice-only corridors provide an envelope of
\emph{where} and \emph{when} an Arctic route might be attractive at the basin
scale, while (ii) realistic planning for specific vessels and ports will
require higher-resolution bathymetry and sea-ice information, which we
address in subsequent work.

\subsubsection{Summary of key implications for large-scale Arctic route diagnostics}
\label{sec:results-summary-implications}

The experiments in Sections~\ref{sec:results_bathy_baseline}–\ref{sec:results_joint}
allow a number of robust conclusions about large-scale Arctic route diagnostics on a
fixed geographic grid.
To provide context for these conclusions, Table~\ref{tab:route_scenarios}
summarises the canonical trans-Arctic route distance and inflation factor under
the main constraint combinations considered in this study.

\begin{table}[htbp]
  \centering
  \caption{Canonical trans-Arctic route distances and inflation factors under
  different constraint sets. $D_{\mathrm{GC}}$ is the great-circle distance
  between origin and destination; $D_{\mathrm{route}}$ is the A*-derived path
  length on the routing grid; $\rho = D_{\mathrm{route}}/D_{\mathrm{GC}}$.}
  \label{tab:route_scenarios}
  \resizebox{\linewidth}{!}{%
  \begin{tabular}{lccc}
    \hline
    Scenario &
    $D_{\mathrm{route}}$ [NM] &
    $\rho$ &
    Notes \\
    \hline
    Great-circle (unconstrained) &
    4149.6 & 1.000 &
    Theoretical lower bound \\
    Sea-only (GEBCO, static) &
    4561.1 & 1.099 &
    Land mask only; no ice or depth constraints \\
    Ice-only, 15 August 2018 ($\mathrm{SIC}<15\%$) &
    5206.0 & 1.255 &
    First date with continuous ice-safe corridor \\
    Ice-only, 15 September 2018 ($\mathrm{SIC}<15\%$) &
    5019.9 & 1.210 &
    Late-summer minimum of ice-constrained distance \\
    Joint depth+ice, 15 September 2018
    ($h_{\min}=20$~m, $\mathrm{SIC}<15\%$) &
    -- & -- &
    No continuous joint-safe route \\
    \hline
  \end{tabular}%
  }
\end{table}

First, enforcing sea-only feasibility on GEBCO~2024 already lengthens the canonical
trans-Arctic connection by about $10\%$ relative to a great-circle, even before
any environmental constraints are applied.  This confirms that widely used
rule-of-thumb shortcuts for trans-Arctic corridors underestimate path length once
sharp coastlines and islands are honoured on a realistic bathymetric grid.

Second, depth constraints at $0.5^{\circ}$ resolution are relatively benign for
open-ocean routing.  For representative under-keel limits of
$h_{\min}=20$--$50$~m, more than $85\%$ of the sea mask remains
depth-safe, and a sea-only route persists for $h_{\min}=20$~m.
Bathymetry therefore acts primarily as a static structural constraint, shaping
the corridor but not fundamentally blocking the basin-scale connection at this
resolution.

Third, historical sea-ice concentration is the dominant time-varying control on
trans-Arctic connectivity in the present framework.  The fraction of ice-safe
sea (SIC~$<15\%$) increases from roughly $60\%$ in mid-June to over $80\%$ in
mid-September, and a continuous trans-Arctic ice corridor only appears from
mid-August onwards.  Once open, the resulting ice-constrained routes are
systematically longer than the sea-only benchmark, but their excess distance
decreases by about 200 NM between August and late September as the
marginal ice zone retreats.

Fourth, at the large scales considered here the existence and length of an
ice-constrained route are only weakly sensitive to the exact SIC threshold in
the range $5$--$50\%$.  Changing the threshold modifies the fraction of
ice-safe sea by a few percentage points but only perturbs route length by
$\sim 20$–$30$~NM when a route exists.  For basin-scale diagnostics this suggests
that the choice of SIC threshold is less critical than the choice of season and
year, although it will matter more once finer grids and coastal passages are
resolved.

Finally, combining bathymetry and ice into a joint feasibility mask reveals that
even a relatively mild under-keel constraint ($h_{\min}=20$~m) can fragment the
apparently open late-summer ice corridor.  On 15~September~2018, only
$74.6\%$ of sea cells satisfy both depth and SIC criteria simultaneously, split
into $35$ disconnected components, and the canonical origin and destination fall
into different components.  At $0.5^{\circ}$ resolution there is therefore no
continuous route that is simultaneously sea-only, depth-safe and ice-safe along
its full length.

Taken together, these results position the present study as a basin-scale,
\emph{offline} screening tool: it quantifies how static bathymetry and
historical ice fields shape the feasibility and excess distance of trans-Arctic
routes, and delineates the temporal window during which such connections are
even theoretically possible.  In the companion work we build on this foundation
by moving to time-evolving forecasts, finer grids and multi-objective cost
functions better aligned with operational routing practice.
\section{Conclusions}
\label{sec:conclusions}

This study developed an offline, grid-based framework for diagnosing the
feasibility and geometry of trans-Arctic shipping routes under static
bathymetric and historical sea-ice constraints.  Using GEBCO~2024
bathymetry regridded to a $0.5^{\circ}$ pan-Arctic corridor
($40$--$85^{\circ}$N, $20^{\circ}$W–$180^{\circ}$E), we constructed a
sea-only routing mask and applied A* search to a canonical west–east
origin–destination pair.  Even without environmental forcing, enforcing
strict sea-only feasibility increased route length from $4{,}150$~NM
(great-circle) to about $4{,}560$~NM (sea-only), i.e.\ a $\sim 10\%$
penalty arising purely from the shape of coastlines and islands on a
realistic bathymetric grid.

Static depth constraints at $0.5^{\circ}$ resolution were found to be
relatively benign at basin scale.  For representative under-keel limits
of $h_{\min}=20$–$50$~m, more than $85\%$ of the sea mask remained
depth-safe, and a sea-only route persisted for $h_{\min}=20$~m.
Depth therefore acts as a structural constraint---eliminating shallow
shelves and passages---but does not, by itself, prevent large-scale
trans-Arctic connectivity in this configuration.  Bathymetry still
matters, however, for shaping where such routes can run and for
interpreting how close candidate paths approach shallow and coastal
zones.

Historical sea-ice concentration emerged as the dominant
time-varying control.  Using summer 2018 CMEMS sea-ice concentration
regridded to the same $0.5^{\circ}$ routing grid and an ice-safety
threshold of $\mathrm{SIC}<15\%$, the fraction of ice-safe sea increased
from roughly $60\%$ in mid-June to over $80\%$ in mid-September.
Continuous ice-safe trans-Arctic routes appeared only from mid-August
onwards and were consistently longer than the sea-only benchmark:
$\sim 5{,}200$~NM in mid-August, decreasing to $\sim 5{,}000$~NM by
late September as the marginal ice zone retreated.  The excess-distance
penalty relative to great-circle therefore drops from about $25\%$ to
$20\%$ over this late-summer window, highlighting how the timing of
Arctic transits trades off against both feasibility and path length.

At the large scales considered here, route geometry was only weakly
sensitive to the precise sea-ice threshold once a route existed.
Varying the concentration threshold between $5$ and $50\%$ changed the
fraction of ice-safe sea by a few percentage points and altered
route length by only 20–30 NM.  For coarse-grid,
basin-scale diagnostics this suggests that the choice of SIC threshold
is less critical than the choice of season and year, although the
threshold will become more important once narrower passages and coastal
traffic lanes are resolved.

Combining bathymetry and ice into a joint feasibility mask exposed a
more restrictive picture.  For a relatively mild depth constraint
($h_{\min}=20$~m) and $\mathrm{SIC}<15\%$ on 15~September~2018, only
about $75\%$ of sea cells were simultaneously depth-safe and ice-safe,
split into 35 disconnected components.  In this joint mask the canonical
origin and destination fall into different components, and no fully
depth- and ice-safe trans-Arctic route exists at $0.5^{\circ}$
resolution.  This underlines that optimistic statements about a
``seasonally ice-free'' Arctic can overstate actual navigable
connectivity once realistic draft constraints are imposed.

The present framework has several limitations that also point directly
to future work.  First, the routing grid is coarse
($0.5^{\circ}\approx 55$~km), so narrow straits, detailed coastline
geometry and port approaches are not resolved; the results are intended
as basin-scale diagnostics, not operational tracks.  Second, only a
single canonical origin–destination pair and a single summer season
(2018) were analysed; extending the analysis to a portfolio of
Europe–Asia and intra-Arctic ODs and to multi-year sea-ice statistics
would better capture inter-annual variability.  Third, cost was defined
purely as geographic distance, without explicit representation of
metocean forcing, resistance, fuel consumption, emissions, or risk.
Finally, several safety-relevant factors---sea-ice thickness, ridging,
traffic separation schemes, and region-specific regulatory corridors---are
omitted and should be incorporated before any operational use.

Despite these limitations, the study provides a physically grounded
baseline for large-scale, safety-constrained Arctic routing.  It
quantifies how static bathymetry and historical sea ice jointly shape
the feasibility, timing and excess distance of trans-Arctic connections,
and it delineates when such routes are even theoretically open.
Follow-on work will build directly on this foundation by (i) increasing
spatial resolution and adding multiple ODs, (ii) integrating
time-evolving sea-ice forecasts and metocean forcing into dynamic
cost fields, and (iii) extending the single-distance metric towards
multi-objective formulations that link routing more directly to
fuel consumption, emissions and schedule reliability.
\section*{CRediT authorship contribution statement}
\textbf{Abdella Mohamed:} Conceptualization, Methodology, Coding, Data curation, Writing, Visualization, Investigation, Validation. 
\textbf{Xiangyu Hu:} Supervision, Conceptualization, Methodology, Manuscript review and editing. 
\section*{Acknowledgements}
This work was supported by Technical University of Munich and Everllence (formerly: MAN Energy Solutions).

\section*{Data and Code Availability}

Bathymetry data are taken from the GEBCO~2024 global grid
\citep{GEBCO2024}, available at \url{https://www.gebco.net/}.
Sea-ice concentration fields are taken from the CMEMS Arctic Ocean Sea Ice
Reanalysis \citep{CMEMS_SeaIce_Reanalysis}, available via doi:10.48670/mds-00336.

The Python code used to construct feasibility masks and perform A* routing,
together with configuration files specifying all parameters used in this
study, is openly available at \url{https://doi.org/10.5281/zenodo.17863887}, under an open-source licence.

\section*{Declaration of Generative AI and AI-assisted Technologies in the Preparation of this Work}
The authors used OpenAI ChatGPT to assist with language editing during manuscript preparation. The authors reviewed and edited all AI-generated content and take full responsibility for the scientific integrity and accuracy of the manuscript.

\listoftables
\listoffigures
\clearpage

\appendix
\section*{Appendix A. Supplementary figures}
\setcounter{figure}{0}
\renewcommand{\thefigure}{A\arabic{figure}}

\begin{figure}[htbp]
  \centering
  \includegraphics[width=\linewidth]{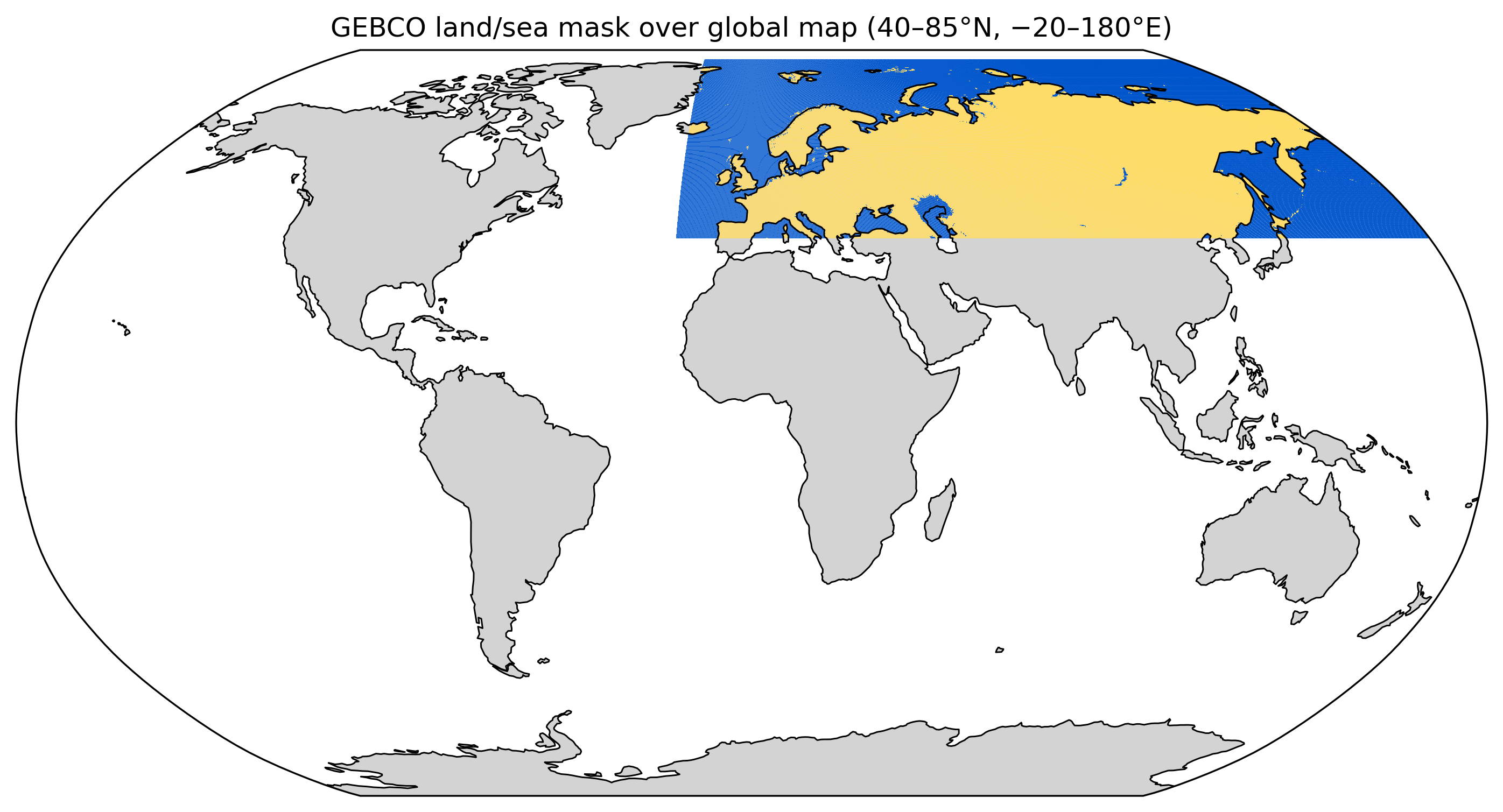}
  \caption{%
  GEBCO~2024 land/sea mask highlighting the Arctic routing domain
  (40--85$^{\circ}$N, 20$^{\circ}$W--180$^{\circ}$E) in the global context.%
  }
  \label{figA_global_domain}
\end{figure}

\begin{figure}[htbp]
  \centering
  \includegraphics[width=\linewidth]{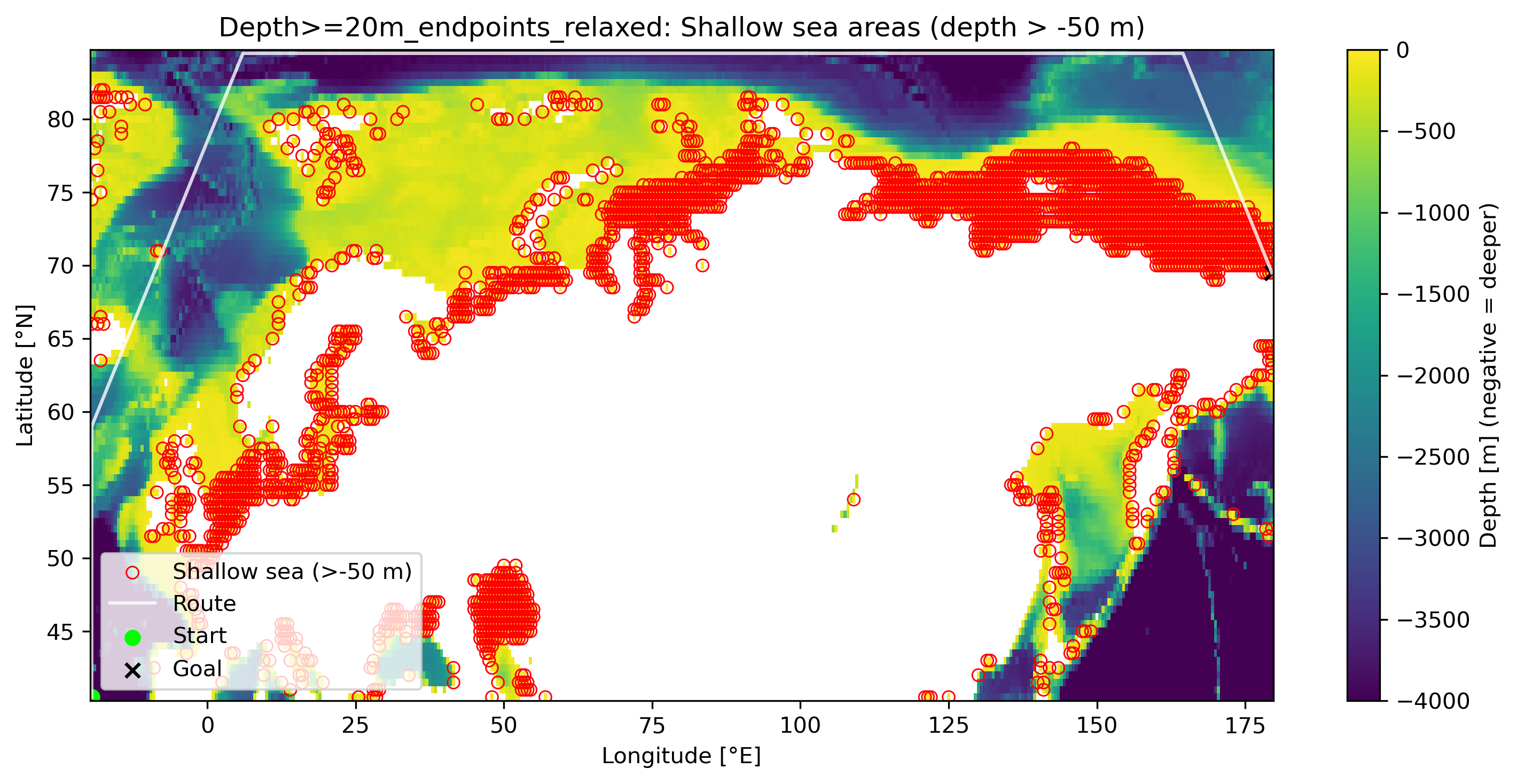}
  \caption{%
  Shallow sea areas shallower than 50~m (red circles) over GEBCO~2024
  bathymetry on the $0.5^{\circ}$ routing grid, together with the
  canonical trans-Arctic sea-only route.%
  }
  \label{figA_shallow_sea}
\end{figure}

\begin{figure}[htbp]
  \centering
  \includegraphics[width=0.8\linewidth]{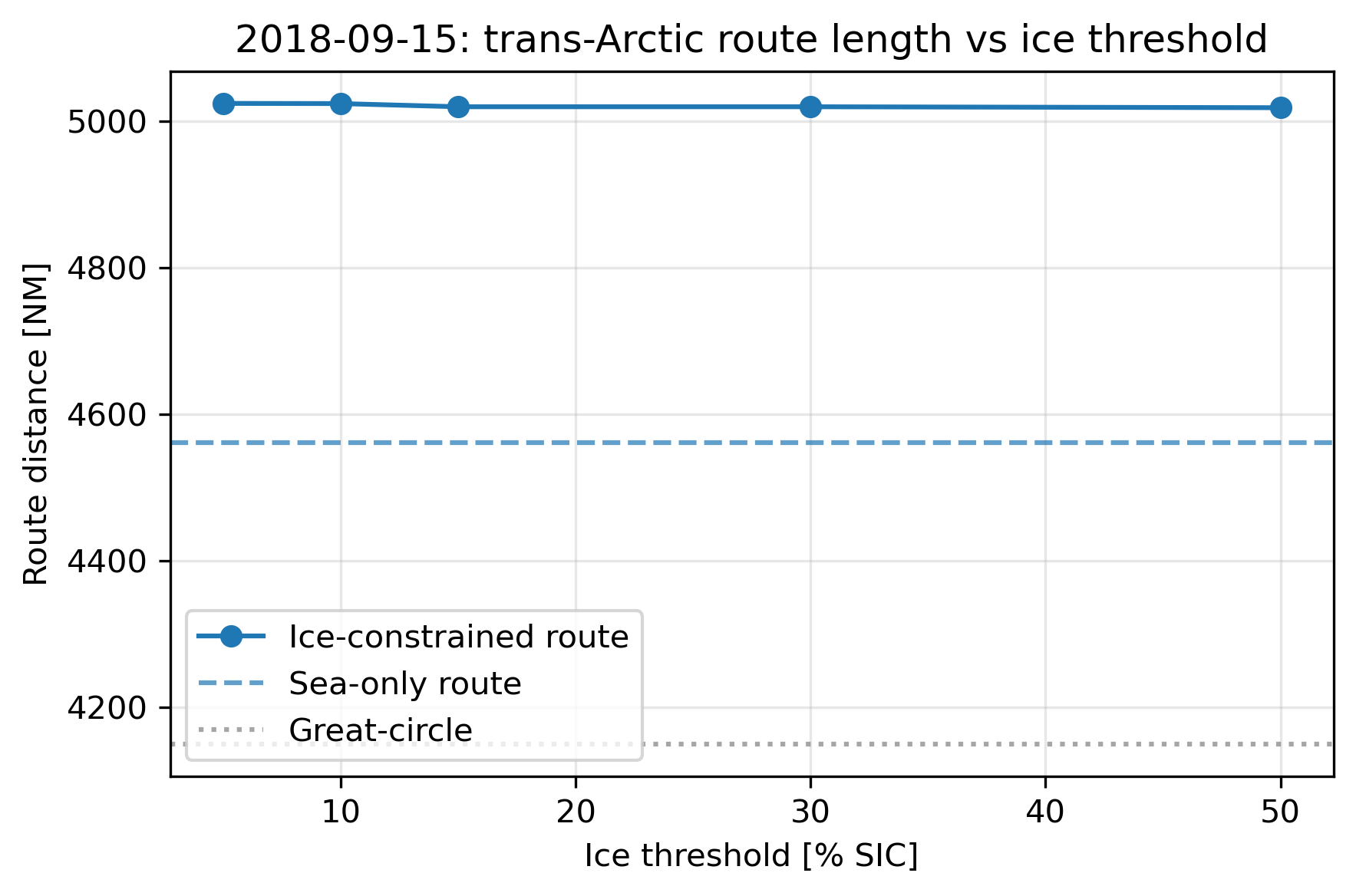}
  \caption{%
  Sensitivity of late-summer (15~September~2018) trans-Arctic route
  distance to the choice of ice-safety threshold. Solid line: ice-constrained
  route; dashed line: bathymetry-only sea route; dotted line: great-circle
  distance.%
  }
  \label{figA_route_vs_ice_threshold}
\end{figure}

\clearpage

\bibliography{main}
\newpage
\end{document}